%% file: main.tex
\documentclass{article}

\usepackage[preprint]{neurips_2026}

\usepackage{microtype}
\usepackage{hyperref}
\usepackage{url}
\usepackage{booktabs}
\usepackage{wrapfig}
\usepackage{colortbl}
\usepackage{graphicx}
\usepackage[utf8]{inputenc} 
\usepackage[T1]{fontenc}    
\usepackage{hyperref}       
\usepackage{url}            
\usepackage{booktabs}       
\usepackage{amsfonts}       
\usepackage{nicefrac}       
\usepackage{microtype}      
\usepackage{xcolor}         
\usepackage{multirow}
\usepackage{enumitem}
\usepackage{bm}%
\usepackage{amsmath}%
\usepackage{amssymb}%
\usepackage{amsthm}
\usepackage{colonequals}
\usepackage{psfrag}
\usepackage{mathrsfs}
\usepackage[most]{tcolorbox}
\usepackage{listings}
\usepackage{xcolor}

\lstdefinestyle{jsonbox}{
  basicstyle=\ttfamily\footnotesize,
  breaklines=true,
  breakatwhitespace=true,
  columns=fullflexible,
  keepspaces=true,
  moredelim=**[is][\color{teal}]{\\key\{}{\}},
  showstringspaces=false,
  frame=none
}

\newtcblisting{promptbox}{
  enhanced,
  breakable,
  colback=gray!10,
  colframe=gray!50,
  boxrule=0.4pt,
  arc=2pt,
  left=4pt,
  right=4pt,
  top=0pt,
  bottom=0pt,
  boxsep=0pt,
  before skip=0pt,
  after skip=0pt,
  width=\linewidth,
  listing only,
  title=,
  listing options={
    style=jsonbox,
    breakindent=0pt,
    breakautoindent=false,
    prebreak=,
    postbreak=
  }
}

\usepackage{siunitx}

\usepackage{algorithm}
\usepackage{algpseudocode}

\usepackage{comment}

\usepackage{lineno}
\definecolor{darkblue}{rgb}{0, 0, 0.5}
\hypersetup{colorlinks=true, citecolor=darkblue, linkcolor=darkblue, urlcolor=darkblue}

\input{notation.tex}

\title{Turning Intent into Specifications:\\ A Benchmark and an Interactive User-Assistant Agent}

\author{%
Hao Wang$^{1,2}$, Ligong Han$^{1,2}$, Kai Xu$^{1,2}$, Akash Srivastava$^{2,3}$\\
$^1$Red Hat AI Innovation\ \ \ 
$^2$MIT-IBM Watson AI Lab\ \ \ 
$^3$IBM Core AI\\
Correspondence to: \texttt{hao-wang@redhat.com}\\[0.5em]
\href{https://huggingface.co/datasets/haowang94/specbench}{\includegraphics[height=2em]{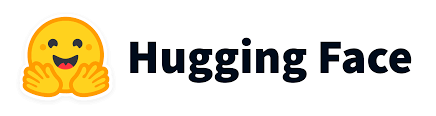}} 
\quad
\href{https://github.com/haowang94/intent2spec}{\includegraphics[height=2em]{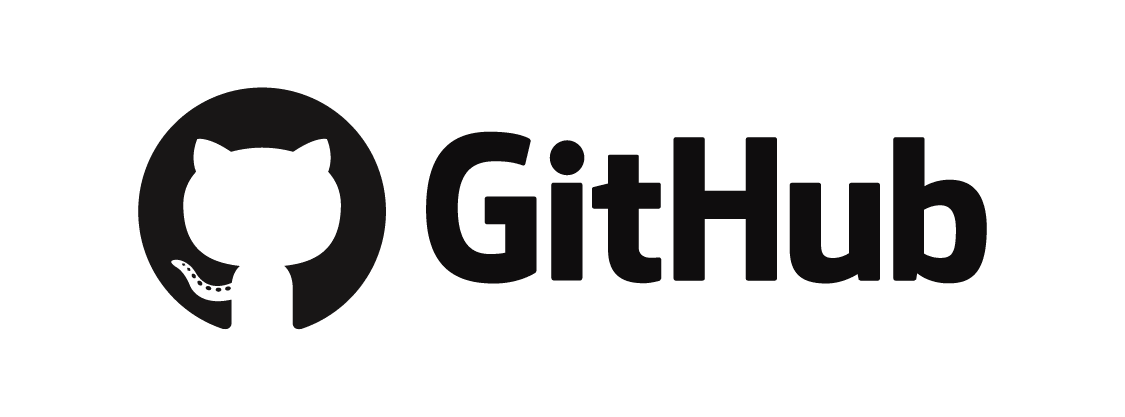}} 
\ \ 
\href{https://haowang94.github.io/blog/specbench/}{\includegraphics[height=2em]{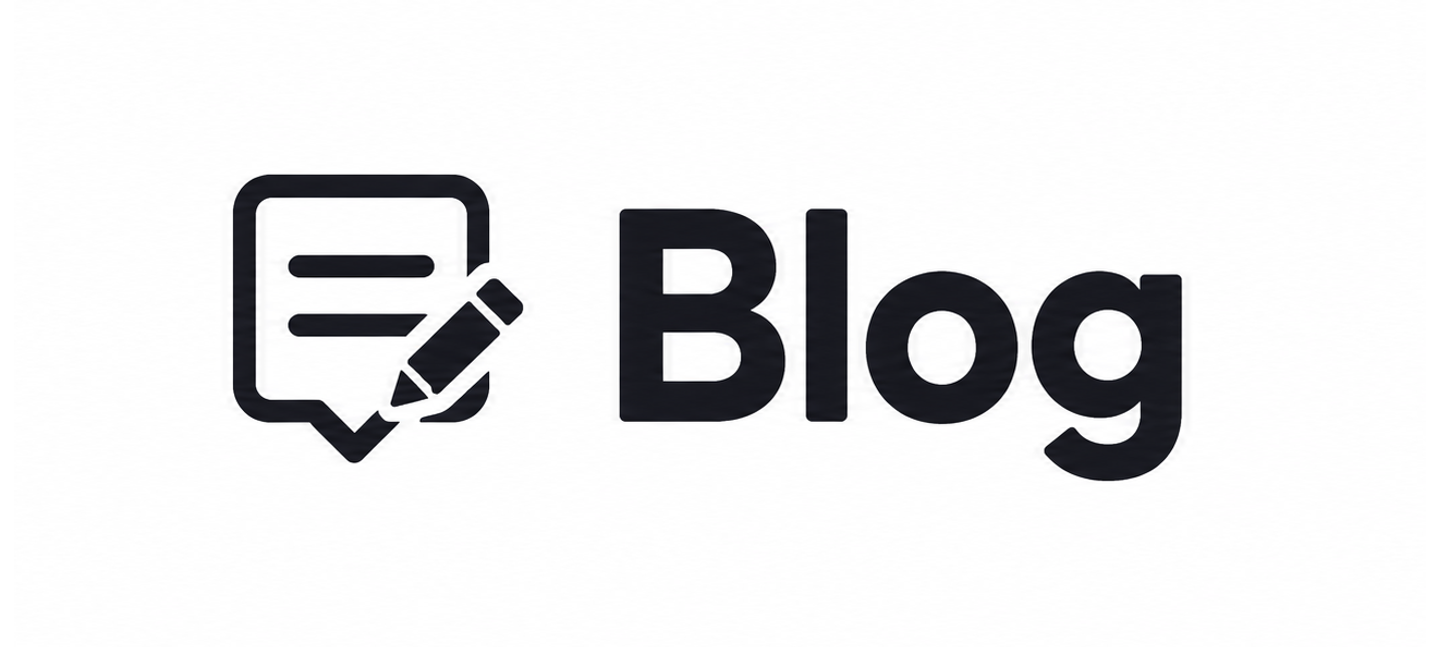}} 
}

\begin{document}

\maketitle

\vspace{-2em}

\begin{abstract}
Today's agents are highly effective at implementing well-scoped software design plans, but user intent is often vague and admits multiple equally valid solutions. In this paper, we introduce \texttt{SpecBench}, a new benchmark for evaluating an agent's ability to translate user intent into a structured, executable specification that aligns with user preferences. The agent is given access to past user conversations and may interact with the user for a fixed number of rounds to ask clarifying questions. We find that existing agents exhibit two extreme behaviors: they either (i) struggle to collaborate proactively with users, entering implementation mode too quickly while overestimating their understanding of user preferences, or (ii) exhaust their question budget by asking about every ambiguous design choice. To address this limitation, we introduce a user-assistant agent: \texttt{Buddy}. It follows a workflow inspired by classical morphological analysis, decomposing user intent into a structured space of design dimensions and candidate choices. It then creates simulated users to evaluate these choices, before engaging the real user to resolve remaining ambiguities and finalize the specification. By shifting the focus from execution to specification, \texttt{SpecBench} and \texttt{Buddy} emphasize agent-user collaboration (not just code generation) as a key frontier in future agent design.
\end{abstract}

\input{section.tex}

\clearpage
\bibliography{references}
\bibliographystyle{alpha}

\clearpage
\appendix

\input{appendix}

\end{document}

%% file: notation.tex
\theoremstyle{definition}

\newcommand{\key}[1]{\textcolor{teal}{#1}}
\tcbset{
    colback=gray!5,
    colframe=black!70,
    arc=6pt,
    boxrule=0.8pt,
    left=10pt,
    right=10pt,
    top=8pt,
    bottom=8pt
}

%% file: section.tex
\section{Introduction}

We are entering an era in which agents can automate a wide range of tasks for individuals, yet they still struggle to understand what users actually want before deciding what to build. This challenge is especially pronounced when a user's initial request is underspecified. For example, a request to ``build a data storage platform'' could imply very different systems: a privacy-conscious user might expect local encrypted storage, while a convenience-oriented user might prefer cloud synchronization. Traditional software development handles this ambiguity through a structured process. A client starts with a business need, a product manager then works with the client to produce a spec sheet that clarifies goals, constraints, success metrics, and requirements. Then engineers review the spec and propose an implementation plan. 
This process is slow, but it exists because software is not just code; it is a set of decisions. AI coding agents can now write functions and fix bugs, but before they start coding, they should help users discover what they want to build by asking clarifying questions and transforming vague ideas into executable specifications.

\begin{figure*}[t]
    \centering
    \includegraphics[width=0.95\linewidth]{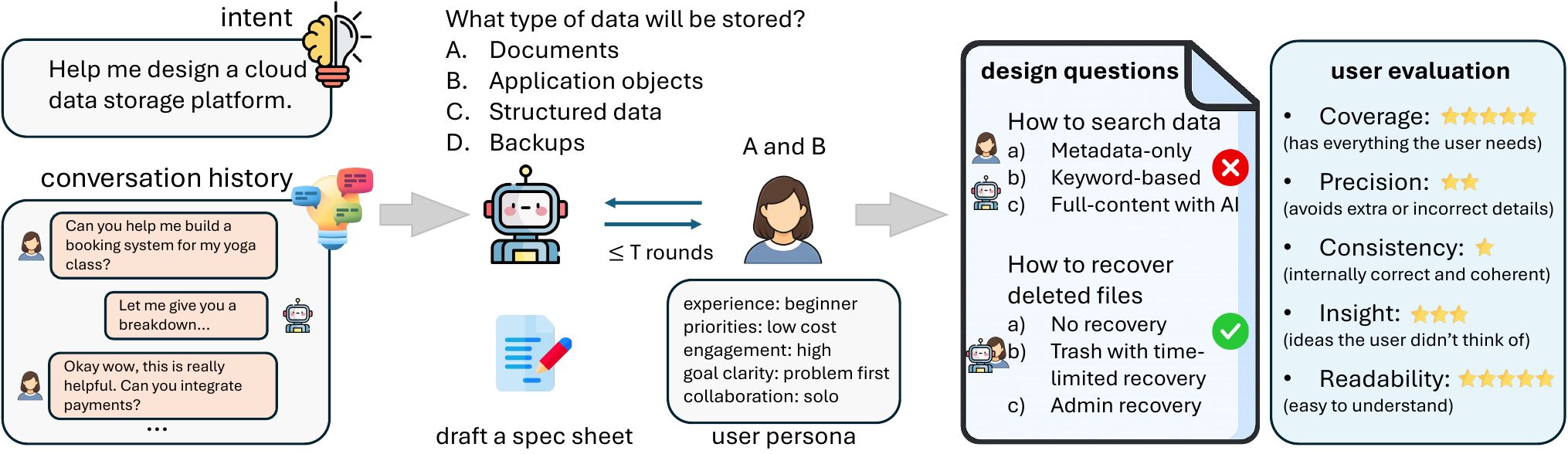}
    \caption{Pipeline of \texttt{SpecBench}. The agent receives a user intent and aims to produce a spec sheet that can be used for implementation. It has access to the user's prior conversation history (from a separate session) and can interact with the user for a fixed number of rounds to ask clarification questions. Task 1: the agent answers a set of design questions, each representing alternative choices for implementing the user's intent. We evaluate performance by comparing the agent's answers with the user's. Task 2: the agent and the user collaboratively write a spec sheet. We evaluate performance using an LLM-as-judge framework that assesses the quality of the spec sheet across five dimensions.
    }
    \label{fig:benchmark}
\end{figure*}

In this paper, we introduce a new benchmark, \texttt{SpecBench} (Figure~\ref{fig:benchmark}). It is designed to evaluate an agent's ability to translate a user's vague intent into a structured spec sheet that faithfully aligns with the user's underlying preferences. 
The benchmark consists of 50 software design tasks across diverse domains (e.g., personal use, education, and entertainment) and incorporates 48 simulated user personas. 
For each instance, the agent is provided with an initial intent (a brief software design request) along with the user's past conversation history. The agent is allowed to engage in a fixed number of interaction rounds, during which it may ask clarifying questions before producing a final structured spec sheet. 
We evaluate agent performance using two complementary approaches. First, we assess alignment by asking both the agent and the simulated user to independently answer a set of design questions, and measure the discrepancy between their responses. Second, we adopt an LLM-as-judge evaluation, where a panel of evaluation models, rate the spec sheet on five criteria: coverage, precision, internal consistency, insight, and readability (each scored 1 to 5 with level-anchored rubrics). 
We evaluate state-of-the-art LLM-based agents: \texttt{Claude Code}, \texttt{Gemini CLI}, and \texttt{Cursor CLI}.  Our main observations are:
\begin{itemize}[leftmargin=1em]
\item \textbf{Some CLI agents tend to behave more like solo workers than collaborative peers.} LLM-based agents, specifically \texttt{Gemini CLI}, tend to be less proactive in collaborating with users. Even when explicitly instructed to continue interviewing the user for a fixed number of rounds and to stop early only if they are sufficiently confident in predicting the user's preferences across all design decisions, \texttt{Gemini CLI} still ends the interaction early and misinterprets the user's preferences for many design questions.

\item \textbf{Agents lack effective strategies for querying users.} \texttt{Claude Code} and \texttt{Cursor CLI} tend to keep asking questions whenever any ambiguity remains, often consuming all available opportunities to query the user. However, the benefit of additional queries is limited, and in some cases, more interactions even reduce the quality of the resulting spec sheet.
\end{itemize}

To address these limitations, we introduce \texttt{Buddy}, an interactive user-assistant agent for drafting spec sheets from user intent. \texttt{Buddy}'s core innovation is to resolve two central challenges in specification drafting: identifying the most important dimensions that require user clarification, and minimizing user interaction while ensuring that the final spec sheet remains aligned with the user's preferences.

\texttt{Buddy} is inspired by classical morphological analysis \citep{zwicky1967morphological}. It decomposes a spec sheet into independent design dimensions, each corresponding to a distinct preference axis, such as feature scope or user experience style, and explores alternatives along each dimension. The agent then analyzes the user's prior conversations with coding agents to construct two complementary forms of memory: a user profile, which captures latent attributes such as technical expertise and risk tolerance, and a conversation summary, which represents observed behavioral patterns across previous projects. The profile captures who the user is, while the summary captures how the user tends to behave.
Together, these memory representations serve as the basis for creating simulated users, which evaluate and select among the design alternatives for each dimension. When the simulated users agree on a choice with high confidence, \texttt{Buddy} adopts that decision without querying the real user. For any remaining ambiguous dimensions, the user interaction module generates candidate clarification questions, selects the question expected to most effectively reduce uncertainty, and poses it to the user. Finally, \texttt{Buddy} verifies and reviews the synthesized spec sheet by checking for internal consistency across sections, alignment with the original project proposal, and faithful adherence to the user's explicitly stated requirements.

\subsection{Related Work}

\paragraph{Uncertainty quantification.}
We observe that LLM-based agents often think they have gathered sufficient information about a user's intent and generate a spec sheet, yet still make choices that fail to align with user preferences. This behavior is consistent with prior work showing that LLMs tend to exhibit overconfidence \citep{achiam2023gpt,zhu2023calibration,tian2023just,lin2022teaching}. This line of work focuses on confidence in a model's final answer (for example, in mathematical reasoning or knowledge-based question answering) where uncertainty arises from incomplete knowledge or limited reasoning ability \citep{damani2025beyond,park2025know,chuang2023dola}. In contrast, we examine a different source of uncertainty: ambiguity in the user's initial intent, where multiple valid implementation plans may exist. In this regard, our work is closely related to prior research \citep{min2020ambigqa,saeidi2018interpretation,deng2025interactcomp,rahmani2023survey,andukuri2024star,rajpurkar2018know,kirichenko2025abstentionbench,song2025hallucination,kalai2025language,vijayvargiya2026ambig}. Unlike them, we study how agents can leverage a user's conversation history to reduce such ambiguity, and how they can proactively interact with users to further clarify and resolve it.

\paragraph{Agent memory and personalized agent.}
A growing body of work studies how agent memory can organize user interactions over time to improve future performance and personalization \citep{zhong2024memorybank,pan2025memory,tan2025prospect,maharana2024evaluating,wu2024longmemeval,singh2024personal,pan2025memory,hu2025evaluating}. Broadly, they aim to enable agents to accumulate experience, adapt to user preferences, and maintain a long-term memory across sessions. For example, \citep{xu2025mem} propose A-Mem that autonomously builds and updates structured long-term memory to improve performance over time; \citep{chhikara2025mem0} introduce structured memory management for personalization; \citep{zhang2025agentic} leverage evolving playbooks to manage context; and \citep{zhong2024memorybank} propose a memory bank composed of event summaries and user portraits. Additionally, GSD\footnote{https://github.com/gsd-build/get-shit-done} also turns project ideas into structured specs and maintains a user profile but it only changes how the agents' question are presented to the user and it never makes decision for the user. Compared with them, we do not focus on designing a general-purpose memory architecture. Instead, we study how memory can be leveraged to guide agents in making software design decisions that align with user preferences. Moreover, rather than focusing solely on improving agent-user interaction, we introduce simulated users that evolves over time: starting with limited knowledge about the user and requiring proactive engagement, and gradually developing a comprehensive user profile that enables the agent to autonomously make many design decisions on the user's behalf.

\paragraph{Human-in-the-loop.} Our work is also related to research on human-AI collaboration \citep{wang2025adaptive,feng2024large,gao2024enhancing,lu2023does,li2023eliciting,qiu2026bayesian,qian2025userbench,edwards2026ask}. For example, \citep{wu2025collabllm} introduce a training recipe for improving multi-turn human-LLM collaboration; \citep{mysore2025prototypical} examine patterns of user-LLM collaborative behavior in English writing sessions; \citep{mu2023clarifygpt} study ambiguous software requirements by detecting ambiguity, asking clarification questions, and refining code-generation requirements. 
We do not focus on code generation. Instead, we study an earlier and more open-ended stage of software development: transforming vague user intent into a structured spec sheet. In this setting, the central challenge is not to resolve missing implementation details, but to infer user-specific design preferences and trade-offs. The agent uses the user's past interaction history as memory to infer latent preferences, propose targeted clarification questions for guidance, and ultimately produce a coherent spec sheet aligned with the user's preferences.
We remark that this formulation closely resembles a classical active learning problem \citep{cohn1994improving,fang2017learning}, where the learner selectively queries an oracle to maximize information gain.

\section{\texttt{SpecBench} Overview}

We introduce a benchmark to evaluate an agent's ability to interact with users and progressively refine a vague intent into a structured spec sheet. The agent is initially given a high-level user intent (e.g., ``build a personal website for showcasing my work.'') and can interact with the user for a fixed number of rounds. Through these interactions, the agent gathers information about the user's preferences and eventually produces a structured spec sheet. We highlight some challenges of this benchmark: 
\begin{itemize}[leftmargin=1em]
\item \textbf{No single ground truth.} For the same initial request, different users may have very different preferences. For example, some users may prioritize visual presentation in designing their website, while others may prefer a minimal, fast-loading site. As a result, multiple spec sheets may be equally valid, so the agent must infer the user's preferences and strategically choose which questions to ask. Moreover, user preferences often span multiple dimensions, and knowing a preference along one dimension does not necessarily determine answers to all design questions.

\item \textbf{Accuracy-interaction trade-off.} The agent can interact with the user for only a limited number of rounds, so it must prioritize the most informative questions. At the same time, user feedback may be sparse: users may provide minimal responses, may not fully articulate their preferences, or may lack the domain knowledge needed to answer certain questions meaningfully. For example, when asked about CI/CD pipeline integration, a user might respond, ``I don't think I use any of those things.'' In such cases, the agent spends one round querying the user but receives little actionable signal.

\item \textbf{Users need guidance and can even make mistakes.} When users first submit a request, they often do not have a full picture of the design choices to make and the implementation details involved. The agent's role is therefore not only to elicit preferences, but also to guide users so they can better clarify and articulate their goals. At the same time, the information available to the agent may itself be noisy or misleading. For example, a user's past conversation history may reveal general preferences, but those preferences were expressed in the context of a different project and may not apply to the current one: a user who preferred minimalism for a personal portfolio site may want a feature-rich dashboard for an enterprise tool. Similarly, a user's responses over multiple turns may conflict with one another; they might emphasize simplicity early on, then later request capabilities that introduce substantial complexity. Agents must therefore account for such noise when drafting the final spec sheet.

\end{itemize}

\subsection{Data Generation}

We generate 50 projects (Table~\ref{table:data}) and 48 user persona (Appendix~\ref{sub::persona}). Each project contains an intent (i.e., a brief request specifying what the agent is expected to implement) and a checklist containing 80 single-choice design questions. Each persona is characterized along several dimensions, including technical experience, design priorities, engagement style, goal clarity, and collaboration preference. For each simulated user, we also generate a multi-turn conversation history between the user and a coding assistant discussing general development tasks (e.g., building booking systems, dashboards, or APIs). These conversations serve as a memory that can reveal patterns in the user's preferences, communication style, and decision-making tendencies. Since the original conversations are lengthy, we summarize them by extracting important information and recurring patterns that are valuable for future interactions.

For each project, the user observes the initial request and selects answers from the 80-question checklist that best match their persona's preferences. The user is also allowed to indicate no strong preference, meaning any option is equally acceptable; such questions are excluded from the evaluation. These selections form the ground-truth specification for that user. Notably, most design questions do not have a universally correct answer. The correct choice depends on the user's preference and priorities.

Below, we provide an example project in which a user aims to develop a bias auditing tool for machine learning models. Note that different users' proposals can be different for the same project.

\begin{tcolorbox}[title=\textbf{A User Proposal Describing Their Intent}, colback=blue!5, colframe=blue!60!black]
I want to build a comprehensive ML fairness and bias auditing tool. The idea is to compute standard fairness metrics across protected attributes, support intersectional analysis up to three-way group combinations, and present everything through interactive visualizations that let users drill down from a high-level summary into detailed breakdowns. I want to include bias mitigation across pre-processing, in-processing, and post-processing techniques, with interactive what-if controls so users can see fairness-accuracy trade-offs in real time. On the production side, the tool would continuously monitor deployed models for bias drift and automatically apply interim fixes while alerting humans for longer-term solutions. I also want it to auto-generate model card sections from audit results and support causal fairness analysis beyond just observational metrics. The tool should work for classification, regression, and ranking use cases, and handle both full-access and black-box model auditing gracefully.
\label{ess_1}
\end{tcolorbox}

\subsection{Task and Evaluation}

\paragraph{Task 1: preference elicitation.}
The first task evaluates the agent's ability to infer user preferences and select the most appropriate design choices accordingly. For each project, the agent is provided with an intent, a checklist of design questions, and the user's conversation history. The agent may interact with the user for up to $T$ rounds. In each round, it asks a single multiple-choice question with four options. The user may select any subset of these options or provide a brief free-text response, based on their persona. Once the agent is sufficiently confident, or when it reaches its interaction budget, it submits predictions for all 80 design questions. We evaluate performance by comparing the agent's predictions with the ground truth generated by the user.

Below, we present an example checklist question for the bias auditing tool project. Because the user did not specify a default fairness metric in their proposal (shown above), the agent must either query the user directly or infer the metric from the user's prior conversation history in order to answer the question.

\begin{tcolorbox}[title=\textbf{Checklist Q1. Default Fairness Metric}, colback=green!5, colframe=green!50!black]
\textbf{Which fairness metric should the tool treat as the primary default when auditing a model?}

\begin{enumerate}[leftmargin=1em]
    \item Demographic parity - intuitive and easy to explain to non-technical stakeholders, but ignores base-rate differences across groups
    \item Equalized odds - focuses on error-rate parity which is more legally defensible, but harder to satisfy and explain
    \item Calibration - ensures predicted probabilities are meaningful across groups, but might be incompatible with equalized odds except for perfect classifiers
    \item No default; require users to explicitly select a metric per audit - avoids false comfort but increases setup friction and demands fairness expertise
\end{enumerate}
\end{tcolorbox}

\paragraph{Task 2: collaborative spec sheet drafting.}
The second task evaluates how effectively the agent collaborates with a user to co-create a spec sheet. The agent is not provided with a predefined checklist of design questions and must instead generate a structured spec sheet with 9 sections: project overview, core features, non functional requirements, technical decisions, user experience, out of scope, success metrics, milestones \& priorities, constraints. 
We require the simulated user to first articulate their requirements (i.e., what they want and do not want) for each section. Then these requirements are incorporated into their system prompt and used to ground both their responses and the final evaluation. 
We evaluate the quality of the spec sheet using an LLM-as-judge framework, in which a panel of LLM judges assesses each spec sheet across five criteria: \emph{coverage}, which assesses how well the spec sheet captures the user's requirements; \emph{precision}, which evaluates whether the spec sheet avoids including unwanted or contradictory items; \emph{consistency}, which measures whether the spec sheet is internally coherent and free of contradictions; \emph{insight}, which considers whether the spec sheet introduces useful ideas or considerations that the user had not initially identified but finds valuable; and \emph{readability}, which examines how clear, concise, and well-structured the spec sheet is.

We defer more details of \texttt{SpecBench} to Appendix~\ref{append::benchmark}.

\section{\texttt{Buddy}: A User-Assistant Agent}

We introduce \texttt{Buddy}, an agent that interprets a user's intent and collaborates with them to produce a structured spec sheet. \texttt{Buddy}'s workflow is inspired by classical morphological analysis \citep{zwicky1967morphological, ritchey2018general}, which decomposes a problem into independent dimensions and explores alternative possibilities for each. 
Concretely, \texttt{Buddy} first writes a morphological chart that captures design alternatives corresponding to different user preferences. It then creates simulated users to evaluate and select options that can be directly inferred from the user's preferences. For unresolved decisions, \texttt{Buddy} makes a strategy to query the real user. Finally, it incorporates the user's responses to refine, review, and complete the final spec sheet.

\begin{figure*}[t]
    \centering
    \includegraphics[width=0.93\linewidth]{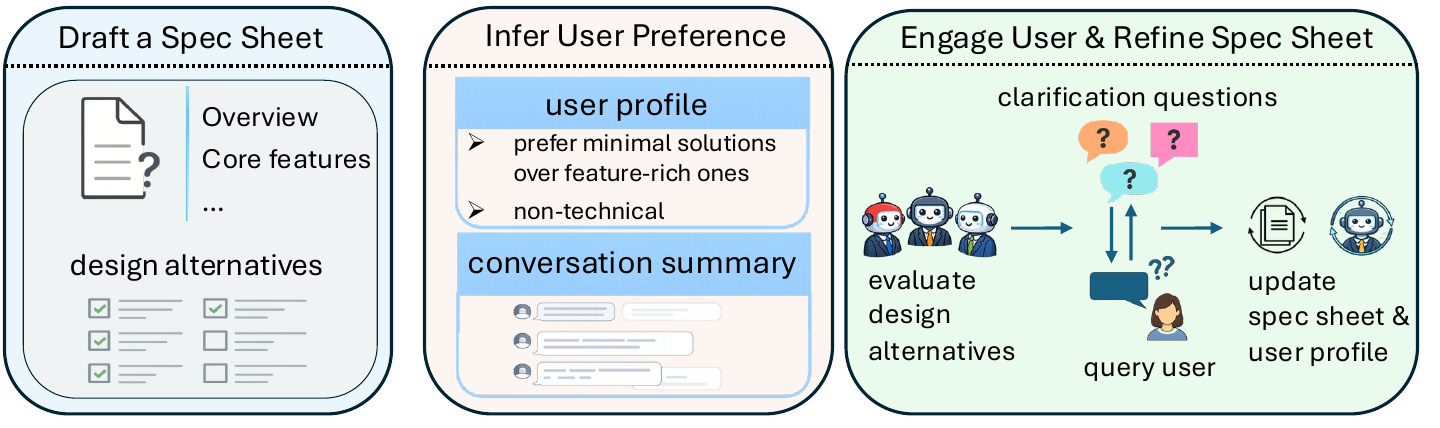}
    \caption{\texttt{Buddy} is an agent that works with the user to create a spec sheet based on their needs. It starts by drafting an initial version and applying morphological analysis to create different design alternatives. Then, it builds a user profile by learning from the past conversations. Next, \texttt{Buddy} suggests clarification questions. Based on the user profile, it creates three simulated users and uses them to determine which questions can be answered directly from the user's inferred preferences. \texttt{Buddy} asks the user the remaining questions, collects their feedback, and then updates and reviews the spec sheet.
    }
    \label{fig:myagent}
\end{figure*}

\paragraph{Draft a spec sheet.} \texttt{Buddy} takes the user's intent and generates a draft spec sheet, together with assumptions (details the user did not specify but the agent introduces to maintain coherence) and open questions (details the agent cannot resolve without user input). It then uses these elements to construct a morphological chart: a structured breakdown of preference dimensions along which different users may make different choices. Each dimension captures a user-facing preference, rather than an implementation choice, and presents exactly four alternatives describing meaningfully different approaches to scope, behavior, or user experience. For example, when building a personal website, one dimension might be privacy level, with options such as: (A) fully public, (B) public with hidden contact information, (C) password-protected, or (D) invite-only access.
Dimensions are restricted to choices grounded in the user's proposal. If the proposal does not mention a capability, no dimension is created for it, preventing the questioning process from expanding the task beyond the user's original proposal.

\paragraph{Infer user preference} To align the spec sheet with user preferences, \texttt{Buddy} analyzes a user's past session history to extract two outputs: a user profile and a conversation summary. The user profile is a structured set of items (similar to playbooks in \citep{zhang2025agentic}), where each item is categorized into: \emph{fact} (e.g., the user is a senior engineer), \emph{soft preference} (e.g., the user prefers simplicity), and \emph{hard constraint} (e.g., security is non-negotiable). Each item is supported with evidences, such as direct quotes from the user's prompts. In addition, \texttt{Buddy} produces a behavioral summary that captures how the user engages over time. It includes recurring interaction patterns, response styles, and the user's priorities. Conceptually, this process can be seen as inferring latent variables (user preferences and behavioral traits) from observed data (conversation history), thereby reducing the influence of irrelevant context.

\paragraph{Engage user and refine spec sheet.} The spec sheet drafted in the first step includes many design alternatives that require user input. However, asking the user to resolve each choice individually would result in excessive interaction. To reduce user effort, we adopt the following approach. First, we introduce three user-simulator sub-agents to make preliminary decisions for each design choice.
These sub-agents all have access to the conversation summary but differ in the amount of user profile information available to them. One has access to the full user profile, another has 20\% of the profile attributes randomly removed, and the third has no profile information. They also receive different system prompt instructions to encourage diverse reasoning. A design choice is finalized only if all three simulators select the same option with high confidence over the alternatives. This approach is inspired by classical MC dropout \citep{gal2016dropout} and serves as a conservative filter to identify decisions that can be reliably inferred from user preferences. 
Afterward, \texttt{Buddy} proposes clarification questions for the remaining unresolved choices and queries the user. The user's responses are then used to iteratively refine and update the spec sheet. Finally, \texttt{Buddy} synthesizes the final specification by integrating all resolved preference dimensions and performs a post-synthesis review to check for missing requirements, unnecessary additions, and internal inconsistencies.

\section{Numerical Experiments}

We compare \texttt{Buddy} with baseline agents on the two tasks in \texttt{SpecBench}, and describe the experimental setup and main observations.

\subsection{Baseline Agents}

We evaluate three LLM agents in \texttt{SpecBench}: \texttt{Claude Code} (\texttt{Opus-4.6}), \texttt{Gemini CLI} (\texttt{Gemini-2.5-pro}), and \texttt{Cursor CLI} (\texttt{GPT-5.4-mini}), using a standard two-stage protocol. In the planning phase, all built-in and MCP tools are disabled. The agent is limited to the task intent, the user's conversation history, and (for Task 1) the checklist. It must produce a strategic plan that (i) identifies ambiguous checklist items or design alternatives for implementing the user's request and (ii) outlines how to interact with the user over $T$ rounds.

\begin{figure*}[t]
    \centering
    \includegraphics[width=0.95\linewidth]{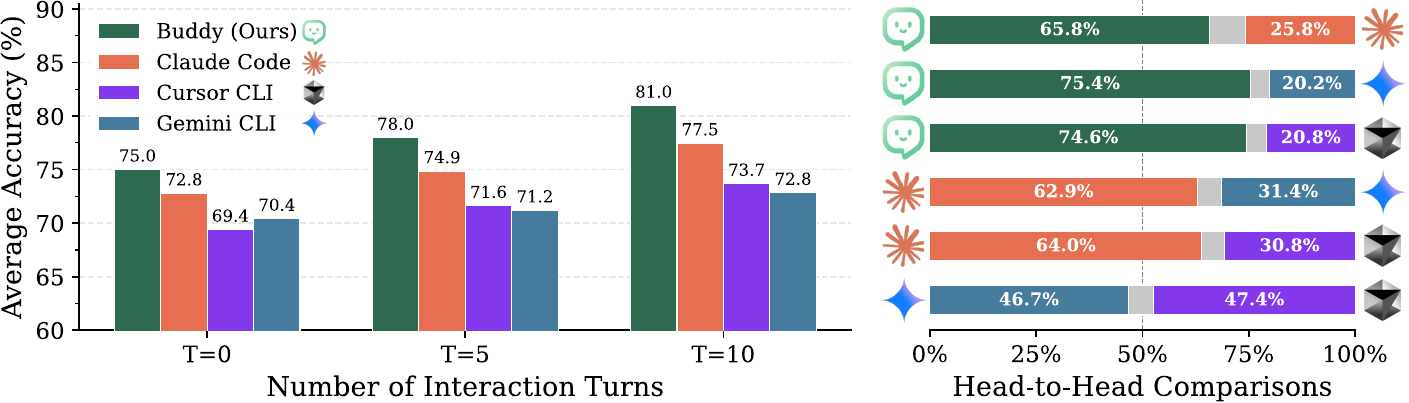}
    \caption{\texttt{SpecBench} Task 1 results. We evaluate \texttt{Claude Code}, \texttt{Gemini CLI}, \texttt{Cursor CLI} (with \texttt{GPT-5.4-mini}), and \texttt{Buddy} using both average accuracy and head-to-head win rates. As shown, \texttt{Buddy} consistently outperforms existing baseline agents both in average score and head-to-head win rate.
    }
    \label{fig:exp_task1}
\end{figure*}

\definecolor{lightblue}{rgb}{0.88, 0.92, 1}
\begin{table*}[t]
\small
\centering
\resizebox{0.92\textwidth}{!}{
\renewcommand{\arraystretch}{1.25}
\begin{tabular}{lcccccc}
\toprule
\textbf{Agent} & \textbf{T0$\rightarrow$T5 $\uparrow$} & \textbf{T5$\rightarrow$T10 $\uparrow$} & \textbf{T0$\rightarrow$T10 $\uparrow$} & \textbf{\# queries (T5)} & \textbf{\# queries (T10)} & \textbf{Early Submit} \\
\midrule
\rowcolor{lightblue}
Buddy & \textbf{74\%} & \textbf{77\%} & \textbf{92\%} & 5.0 & 10.0 & 0\% \\
Claude Code & 62\% & 67\% & 80\% & 5.0 & 10.0 & 1\% \\
\rowcolor{lightblue}
Cursor CLI & 62\% & 61\% & 78\% & 5.0 & 9.9 & 3\% \\
Gemini CLI & 50\% & 57\% & 61\% & 4.9 & 9.2 & 17\% \\
\bottomrule
\end{tabular}
}
\caption{We evaluate how effectively different agents engage with users and how many queries they ask on average in Task~1. Specifically, T0$\rightarrow$T5 $\uparrow$, T5$\rightarrow$T10 $\uparrow$, and T0$\rightarrow$T10 $\uparrow$ report the number of data instances for which the agent score improves when the query budget is increased from 0 to 5, from 5 to 10, and from 0 to 10, respectively. \# queries (T5) and \# queries (T10) report the average number of queries actually asked by the agent under budgets of 5 and 10, respectively. Early submit rate reports the fraction of runs in which the agent stops querying the user before using up the available query budget. We observe that \texttt{Gemini CLI} has high early submit rate. It stops querying the user early and gains little benefit from user interaction.
}
\label{table:agent_scaling}
\end{table*}

In the execution phase, the agent is provided with its plan from the previous stage and given access to two MCP tools: \texttt{query\_user}, which allows it to pose questions to a simulated user (an LLM role-playing the target persona), and \texttt{submit\_predictions}, which accepts the agent's final outputs: either predictions for all 80 checklist questions (Task 1) or a structured spec sheet (Task 2). During the interaction, after each user response, the agent performs a reflection step: it summarizes what it learned from the last answer, updates its understanding of the user's preferences, and determines whether and how to revise its plan based on the new information before posing the next question. We provide their system prompts for reproducibility in Appendix~\ref{append::existing_agents}.

\subsection{Experimental Setup}

For Task 1, we skip the spec-sheet drafting and final spec-sheet review steps (Figure~\ref{fig:myagent}) because the objective is to predict the user's preferences on the checklist questions rather than to produce a complete spec sheet. For a fair comparison, the baselines are likewise required only to produce checklist predictions, without generating a full spec sheet. We use the checklist questions directly as the design alternatives and follow the rest of our pipeline: simulated users select options that can be easily inferred from the user's preferences, and the agent queries the user about the remaining uncertain checklist questions. Finally, \texttt{Buddy} is an LLM-agnostic harness that can be paired with any LLM; in our experiments, we instantiate it with \texttt{Opus-4.6}.

\subsection{Experimental Results}
\begin{figure*}[t]
    \centering
    \includegraphics[width=0.98\linewidth]{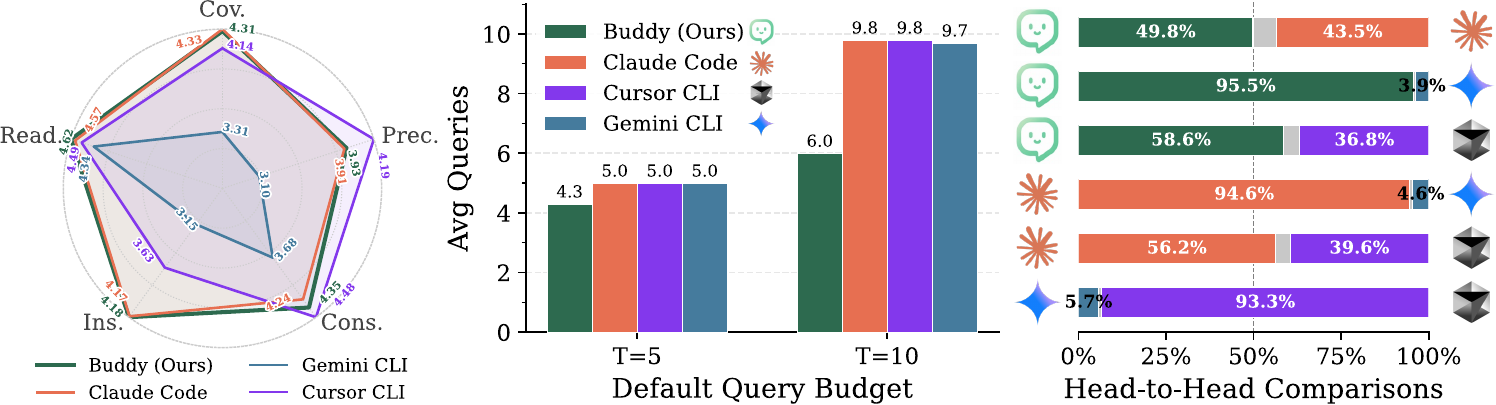}
    \caption{\texttt{SpecBench} Task 2 results. 
    Left: comparison of the spec sheet produced by different agents across five evaluation criteria. The average scores are: \texttt{Buddy} (\textbf{\underline{4.28}}), \texttt{Claude Code} (\underline{4.24}), \texttt{Cursor CLI} (\underline{4.19}), and \texttt{Gemini CLI} (\underline{3.52}). Middle: average number of queries each agent asks the user when given a prior budget of at most $T=5$ or $T=10$ questions. Right: head-to-head win-rate comparison across agents. As shown, \texttt{Buddy} achieves the best overall performance while requiring the fewest user interactions.
    }
    \label{fig:exp_task2}
\end{figure*}

We compare \texttt{Buddy} with baseline agents on two tasks in \texttt{SpecBench}. Figure~\ref{fig:exp_task1} shows the average prediction accuracy on the 80 checklist questions in Task~1 across different interaction turns, along with head-to-head comparisons between agents. Table~\ref{table:agent_scaling} summarizes how much different agents benefit from user interaction and leverage user feedback to better align with user preferences. It also reports the number of queries each agent asks the user, as well as how often an agent terminates the interview and submits predictions for the checklist questions before exhausting the query budget. Figure~\ref{fig:exp_task2} presents the spec sheet quality across different evaluation axes, head-to-head comparisons, and the number of queries each agent asks the user in Task~2.

\subsection{Main Observations}

\paragraph{\texttt{Gemini CLI} behaves like an overconfident solo developer.}
We observe that \texttt{Gemini CLI} has a significantly higher early-submission rate (Table~\ref{table:agent_scaling} last column). In Task 1, we explicitly instruct each agent to stop querying the user only when it is confident enough to predict all 80 checklist questions, or when it has exhausted its query budget. However, \texttt{Gemini CLI} is often overconfident: it stops querying the user early and jumps into execution mode. As a result, both its predictions of the user's checklist preferences and its generated spec sheet underperform those of other baselines.

\paragraph{\texttt{Gemini CLI} does not benefit much from user interaction.}
For example, for many data instances in Task 1, the prediction accuracy on the checklist questions degrades from $T=0$ to $T=5$ and $T=10$ (Table~\ref{table:agent_scaling}, first three columns). In other words, while \texttt{Gemini CLI} is able to infer user preferences effectively from past conversation memory, it struggles to ask effective follow-up questions and leverage the additional interactions to further improve its understanding of user preferences.

We conjecture that the root cause of the above 2 issues is that CLI coding agents are often optimized to complete tasks efficiently. This optimization can make the model prefer taking action (e.g., searching the existing codebase or writing code) over querying the user. In \texttt{SpecBench}, however, asking more questions is valuable: without sufficient user input, the agent must fill information gaps with guesses, and the eventual implementation based on the resulting spec sheet may become a polished version of a wrong software product.

\paragraph{Agents lack effective strategies for querying users.}
Other baseline agents (specifically \texttt{Claude Code} and \texttt{Cursor CLI}) often exhaust the full question budget, querying the user whenever any ambiguity remains and rarely submitting early. This behavior is expected on Task 1: we intentionally include many questions in the checklist, and even with 10 queries it is difficult to elicit the user's preferences across all 80 checklist questions. However, on Task 2, additional conversations with the user appear to yield diminishing returns. Consequently, using all 10 questions becomes a less efficient strategy, suggesting that agents should learn when to stop querying and submit early.

\paragraph{\texttt{Buddy} is more query-efficient and effective at eliciting user preferences.}

Figure~\ref{fig:exp_task2} (Middle) shows that \texttt{Buddy} queries the user significantly fewer times than other agents, while still achieving the highest evaluation scores and outperforming all baselines in head-to-head comparisons. This is because \texttt{Buddy} first uses simulated users to resolve many design decisions, leaving only the most uncertain questions for the actual user. In addition, Figure~\ref{fig:exp_task1} shows that \texttt{Buddy} is substantially better at inferring user preferences from past conversations and incorporating new user responses to further refine its understanding of those preferences.

\section{Conclusion, Limitations, and Societal Impact}
\label{sec::conc_limit}

We introduced \texttt{SpecBench}, a benchmark for evaluating how agents transform vague user intent into user-aligned spec sheets. We also introduced \texttt{Buddy}, an interactive agent that learns user preferences, develops query strategies to resolve ambiguity, and drafts structured spec sheets. Our experiments show that existing CLI-based agents often struggle in this collaborative setting: some end the user interview prematurely and overestimate their understanding of user preferences, while others query the user whenever any ambiguity remains. In contrast, \texttt{Buddy} uses morphological analysis and simulated user agreement to identify which design decisions can be safely inferred and which require direct user input. These findings suggest that effective software-design agents should not merely execute tasks, but should act as collaborative partners that reason about ambiguity, model user-specific trade-offs, and ask targeted questions when uncertainty matters most.

We discuss limitations of this work. First, when testing existing agents, we follow a standard workflow: a planning phase to let agents determine how they will interact with users and then an execution phase where they query the user and draft a spec sheet. We hypothesize that agents' performance can be improved by customizing the workflow and optimizing prompts. However, this would require collecting additional training data from user, and tailoring workflows could demand significant effort. When testing existing agents, we compress the past long conversation history between the user and the coding agent into a summary, inspired by \href{https://reference.langchain.com/python/langchain-classic/memory/summary/ConversationSummaryMemory}{LangChain's} approach to memory. We extract user-specific information that may transfer to future projects. We note that there are many other ways to manage agent memory and hypothesize that more effective methods for evolving user memory might further improve baseline agent performance on \texttt{SpecBench}. Finally, we simulate users with LLMs. We construct detailed user profiles, including traits, backstories, and behavioral patterns, to ground their responses. However, we acknowledge that LLM-simulated users remain fundamentally different from real users. User simulation and digital twins are active areas of research \cite[see e.g.,][]{park2023generative,jia2024can,li2025far}, and we hope future progress in these areas can also benefit \texttt{SpecBench}.

Coding agents are powerful tools that can significantly reduce the time required to develop software. However, they also introduce a barrier: for users with limited software development experience, these tools can be difficult to use. For example, new features emerge constantly, making it unclear what the right workflow is or how to effectively review generated code.
This paper focuses on a specific challenge in addressing this gap: situations where users struggle to clearly articulate their requests. In many cases, user intent is vague and may admit multiple valid implementations. We explore how an agent can guide users in refining such ambiguous ideas into well-structured spec sheets. We introduce both a new benchmark \texttt{SpecBench} and a user-assistant agent \texttt{Buddy} for this goal. We hope our effort can serve as a step towards making coding agents more accessible, collaborative, and effective for a broader range of users.

%% file: appendix.tex
\section{Usage of LLMs}

We use Claude Code to assist in generating benchmark data and implementing our agent. For baseline experiments, we evaluate Claude Code, Cursor CLI (with GPT models), and Gemini CLI. Additionally, we use Opus, GPT, and Gemini models to simulate users in our experiments. We carefully review our codebase and experimental results after using LLMs.

\section{\texttt{SpecBench} Details}
\label{append::benchmark}

\begin{table*}[t]
\small
\centering
\renewcommand{\arraystretch}{1.2}
\begin{tabular}{@{} l l p{0.72\textwidth} @{}}
\toprule
\textbf{Category} & \textbf{Count} & \textbf{Tasks} \\
\midrule

\textbf{Personal} & 10 & personal-autobot (daily task automation bot), personal-blog (blog \& newsletter platform), personal-dataplatform (data collection \& visualization), personal-ecommerce (online storefront), personal-finance (budgeting dashboard), personal-healthtracker (health \& fitness tracker), personal-homeinventory (home inventory for insurance), personal-mealplanner (recipe \& meal planner), personal-mobileapp (expense tracking mobile app), personal-website (portfolio website) \\

\textbf{Startup} & 5 & startup-scheduling (appointment scheduling platform), startup-marketplace (two-sided service marketplace), startup-feedback (customer feedback \& NPS), startup-courses (online course platform), startup-socialtools (social media management) \\

\textbf{Enterprise} & 5 & enterprise-onboarding (employee onboarding portal), enterprise-incidents (incident management system), enterprise-approvals (purchase approval workflows), enterprise-assets (IT asset management), enterprise-knowledge (internal knowledge base) \\

\textbf{Education} & 5 & edu-coding (interactive coding tutorials), edu-flashcards (spaced repetition flashcards), edu-quizzes (quiz \& assessment platform), edu-classroom (virtual classroom), edu-progress (student progress dashboard) \\

\textbf{Research} & 5 & research-experiments (ML experiment tracking), research-annotation (data annotation \& labeling), research-literature (literature review manager), research-participants (participant pool management), research-inventory (lab inventory system) \\

\textbf{ML/Technical} & 5 & ml-syntheticdata (synthetic data generation), ml-serving (model serving platform), ml-features (ML feature store), ml-fairness (bias auditing tool), ml-monitoring (production model monitoring) \\

\textbf{Agent} & 5 & agent-personal (AI productivity assistant), agent-support (tier-1 support agent), agent-intent (intent elicitation framework), agent-coderev (automated code review), agent-workflows (multi-agent workflow platform) \\

\textbf{Entertainment} & 5 & entertainment-gamebuilder (2D game creation tool), entertainment-streaming (watch party platform), entertainment-trivia (multiplayer trivia game), entertainment-musicapp (music discovery \& playlists), entertainment-storyengine (interactive fiction authoring) \\

\textbf{Healthcare} & 5 & healthcare-telehealth (video consultation platform), healthcare-medications (medication tracking \& reminders), healthcare-scheduling (clinic appointment scheduling), healthcare-patientportal (patient health records portal), healthcare-symptoms (symptom checker \& health journal) \\

\bottomrule
\end{tabular}
\caption{Overview of the 50 benchmark projects.}
\label{table:data}
\end{table*}

\subsection{Benchmark Data}

\texttt{SpecBench} consists of 50 projects, each representing a software project drawn from nine domain categories (Table~\ref{table:data}). Each project begins with a base proposal: a first-person description in which a user explains what they want to build (an example is shown below). These proposals do not include any implementation details and, hence, can admit many different realizations. They outline goals and constraints without specifying particular methods or architectures. This setup mirrors a real user's initial request to a coding agent, which is often underspecified and ambiguous. For every project, each of the 48 personas rewrites the base proposal in their own voice, reflecting their individual traits and perspectives.

\begin{tcolorbox}[title=\textbf{Base Proposal}, colback=blue!5, colframe=blue!60!black]
I want to build an ML fairness and bias auditing tool. Our company deploys models for lending decisions, hiring screening, and content recommendations, and we need systematic ways to detect and measure bias before and after deployment. The tool should compute standard fairness metrics (demographic parity, equalized odds, calibration) across protected attributes, with support for intersectional analysis and visual reports that non-technical stakeholders can understand. It should support both pre-deployment auditing (on test sets) and continuous monitoring of production predictions, with optional debiasing techniques like reweighting, threshold adjustment, and adversarial debiasing, and integrate with our existing model training pipelines.
\label{ess}
\end{tcolorbox}

For Task 1, we construct, for each of the 50 projects, a checklist of 80 single-choice design questions (one example question is shown below). They capture key design decisions involved in implementing the user's proposal and often do not admit a ground-truth answer. Instead, we define ground truth by having each persona answer all 80 questions for every project. Simulated users are allowed to select a null option if they have no preference for a given question; such questions are excluded from evaluation.

\begin{tcolorbox}[title=\textbf{Checklist Q74. Privacy Handling}, colback=red!5, colframe=red!60!black]
\textbf{How should the tool handle privacy regulations (GDPR, CCPA) that restrict processing of sensitive demographic data?}

\begin{enumerate}[leftmargin=1em]
    \item Provide data minimization features: compute metrics on-the-fly without persisting protected attributes, deleting sensitive data after computation - compliance-friendly but limits reproducibility and historical analysis
    \item Support anonymized/pseudonymized processing where protected attributes are hashed or tokenized but still allow group-level analysis - enables monitoring while reducing re-identification risk but adds anonymization complexity
    \item Rely on legal basis exemptions for bias monitoring (many jurisdictions allow sensitive data processing for discrimination prevention) and process data normally with appropriate documentation - simplest implementation but requires legal review per jurisdiction
    \item Provide configurable data handling modes per jurisdiction, with GDPR-strict, CCPA-strict, and standard modes - covers all cases but significant configuration and testing burden
\end{enumerate}
\end{tcolorbox}

For Task 2 each project, we ask every simulated user to produce a structured requirements document describing both desired and undesired aspects of the project. These documents are organized into nine sections that mirror the spec sheet template the agent is expected to generate: project overview, core features, non-functional requirements, technical decisions, user experience, out of scope, success metrics, milestones and priorities, and constraints. Each section includes explicit ``want'' and ``do not want'' statements grounded in the persona's traits and backstory. These documents serve both as the ground truth for Task 2 evaluation and as the knowledge base from which the simulated user responds to the agent's queries. An example requirement by a simulated user is provided below.

\begin{tcolorbox}[title=\textbf{An Example Requirement for Core Features}, colback=red!5, colframe=red!60!black]
\textbf{Want}: The main thing is the dashboard with all the visualizations. A user should be able to upload a dataset and a model, pick a few columns like race or gender, and instantly see a bunch of fairness scores and charts. The intersectional analysis sounds really impressive, so I definitely want the ability to combine up to three groups and see the results. The ``what-if'' controls for bias mitigation sound cool too. Like, a slider that shows how accuracy changes as you make the model fairer. I also want it to automatically create a ``model card'' summary you can download.

\textbf{Do not want}:Let's not worry about the ``continuously monitor deployed models'' part for now. That sounds complicated. And the ``causal fairness analysis'' seems like a whole research paper on its own. I want to avoid anything that requires a super deep understanding of statistics or theory that I can't pick up reasonably quickly. Let's stick to the standard, well-known fairness metrics. Also, let's just focus on classification models first, like predicting yes/no. Regression and ranking can come later.
\end{tcolorbox}

\subsection{User Persona}
\label{sub::persona}

To ensure simulated users reflect diverse backgrounds, we define five key dimensions of persona categories:
\begin{itemize}[leftmargin=1em]
    \item \textbf{experience}: beginner, junior engineer, PM/founder, senior engineer, designer
    \item \textbf{priorities}: privacy-first, speed/simplicity, feature-rich, low cost
    \item \textbf{engagement}: passive, moderately engaged, highly engaged
    \item \textbf{goal clarity}: clear target, problem-first, exploratory
    \item \textbf{collaboration}: solo, small group, large group
\end{itemize}
Using these dimensions, we then generate individual user profiles. An example of a user persona is:
\begin{promptbox}
\key{"id"}: "maya_chen",
\key{"name"}: "Maya Chen",
\key{"age"}: 34,
\key{"occupation"}: "Runs a small yoga studio",
\key{"location"}: "Portland, OR",
\key{"traits"}: {
  \key{"experience"}: "beginner",
  \key{"priorities"}: "low_cost",
  \key{"engagement"}: "highly_engaged",
  \key{"goal_clarity"}: "problem_first",
  \key{"collaboration"}: "solo"
},
\key{"backstory"}: "Manages bookings for ~60 regular clients through a shared Google Sheet and a WhatsApp group. It's becoming chaotic with double bookings, missed cancellations. She's heard of tools like Mindbody but thinks they're overpriced for her scale. She has no coding experience but taught herself Canva and Squarespace for her studio's website.",
\key{"behavioral_notes"}: [
  "Enthusiastic, asks lots of follow-up questions, sometimes circling back to earlier topics.",
  "Describes problems through stories (\"Last Tuesday, two people showed up for the same 6pm slot and one of them had driven 30 minutes.\").",
  "Doesn't know technical terms like \"CRUD\" or \"API\" -- \"database\" makes her picture Excel.",
  "Gets nervous about ongoing costs.",
  "Willing to put in time to learn but needs analogies to tools she already uses."
],
\key{"example_phrases"}: [
  "I don't need anything fancy, I just need people to stop double-booking.",
  "Wait, so this would be like... a private website just for my clients?",
  "Is there a free version of that? I really can't spend more than $20/month."
]
\end{promptbox}

\subsection{Spec Sheet Structures}

For Task 2, agents are required to produce a structured specification sheet consisting of nine sections, as described below.
\begin{itemize}[leftmargin=2em]
    \item project overview: high-level summary of the project, its goals, and target users.
    \item core features: core features and functionality the system must provide.
    \item non-functional requirements: performance, scalability, reliability, and other quality attributes.
    \item technical decisions: technology choices, platforms, architecture, integrations, and technical trade-offs.
    \item user experience: UX/UI preferences, accessibility, design philosophy.
    \item out of scope: what is explicitly not included in this version.
    \item success metrics: how success will be measured: KPIs, acceptance criteria, definition of done.
    \item milestones and priorities: MVP scope, phasing, build order, what to tackle first vs. later.
    \item constraints: non-technical constraints: budget, timeline, team size, regulatory or compliance requirements.
\end{itemize}

\subsection{User Simulation}

During benchmark runs, users are simulated by three frontier language models: \texttt{Opus-4.6}, \texttt{GPT-5.4-mini}, and \texttt{Gemini-2.5-pro}. Each model is responsible for one-third of the simulations. Each model role-plays its assigned personas using a system prompt that encodes the persona's traits. The prompts also include either the ground-truth answers and reasoning for the checklist questions (Task 1) or the structured requirements document (Task 2). 
The LLM, when role-playing a user, has a restricted response length determined by the engagement level of its simulated persona: highly engaged personas may use up to 150 tokens per response; moderately engaged personas are limited to 80 tokens; passive personas are limited to 40 tokens. This setup creates a realistic challenge: agents interacting with less engaged users must extract useful signal from brief or minimal replies, while those interacting with highly engaged users must efficiently process richer, potentially verbose responses. The agent must infer the user's engagement style from the conversation history and adapt its strategy accordingly.

\subsection{Evaluation}

\paragraph{Task 1.} The agent is given a project proposal along with the full 80-question design checklist. It may interact with a simulated user for up to $T$ rounds, asking a  multiple-choice question per-round (either drawn from the checklist or an original question). If none of the provided options fit, the user may respond in free text. After exhausting its turn budget (or if it is confident enough), the agent stops engaging with the user and predicts the user's answers to all 80 checklist questions.

We evaluate accuracy by comparing the agent's predictions against the user's responses. Questions for which the user selects null (no preference) are excluded from evaluation.

\paragraph{Task 2.} The agent is again given a project proposal and can query the user for up to $T$ rounds. It then produces a structured spec sheet consisting of nine sections: project overview, core features, non-functional requirements, technical decisions, user experience, out of scope, success metrics, milestones and priorities, and constraints.

The specification is evaluated by a panel of three judges: 
\texttt{Opus-4.6}, \texttt{GPT-5.4-mini}, and \texttt{Gemini-2.5-pro}. Each judge role plays the target persona with access to the full requirements document and evaluates the spec sheet along five dimensions:

\begin{itemize}[leftmargin=1em]
    \item Coverage: how well the specification captures the user's requirements
    \item Precision: whether it avoids contradictions or inclusion of unwanted elements
    \item Internal consistency: absence of contradictions within the specification
    \item Insight: addition of useful details beyond explicitly stated requirements
    \item Readability: clarity, conciseness, and organization
\end{itemize}
Each dimension is rated on a 1--5 scale with detailed rubric anchors. Final scores are computed as the mean across all three judges. Using multiple models helps mitigate single-model evaluation bias.

\subsection{Existing Agent Workflow}
\label{append::existing_agents}

We run the existing agents using a standard two-stage protocol consisting of a planning phase and an execution phase. Below, we provide their system prompts for both Task 1 and Task 2.

We begin with the system prompt for the Task 1 planning phase. In this context, the proposal refers to the user's intent (the user's description of what they aim to build). The questions text consists of 80 checklist questions, and the conversation memory is a summary of the user's prior interactions with a coding agent.

\begin{promptbox}
You are an assistant trying to understand a user's design preferences through a limited number of interactions. Your goal is to select appropriate options from a candidate feature checklist so that your selections align with the user's preferences.

Here is the user's proposal: {proposal}. Based on this proposal, select one choice for each of the {num_questions} questions in the candidate feature checklist below.

{questions_text}

## User's Working Style (from prior projects)

The conversation summary below shows how this user approached PREVIOUS projects. Use it to understand their general preferences, values, and constraints.

{conversation_memory}

## Your task

Analyze the {num_questions} design questions above and produce a strategic questioning plan. You will have {num_turns} rounds to ask the user multiple-choice questions. You may ask questions taken directly from the checklist (with their exact options), or create your own original questions.

First, read through all the design questions carefully and classify each one:
- **Predictable** - questions you can answer confidently from the proposal alone or from general conventions (e.g., questions with an obviously standard answer).
- **Ambiguous** - questions whose answer depends on the specific user's preferences, values, or context, and cannot be reliably guessed.

Then, build a questioning plan for your {num_turns}-round budget. Your goal is to predict as many checklist answers correctly as possible. You may ask checklist questions directly or ask broader questions of your own design - whichever you believe will best improve your predictions. Focus your limited rounds on resolving the ambiguous questions.

Output your plan as structured text. Be concise - list only your questioning strategy and the key ambiguous questions you intend to resolve. Do not repeat or paraphrase the checklist questions.

This is the planning phase only - you have no tools and no access to the user right now. Do not switch to execution mode, do not write as if you have already asked questions. Output only your plan.
\end{promptbox}

\vspace*{5mm}
After planning, the agent moves to the execution phase and we provide its system prompt below. 
\vspace*{1mm}
\begin{promptbox}
You are an assistant trying to understand a user's design preferences through a limited number of interactions. Your goal is to select appropriate options from a candidate feature checklist so that your selections align with the user's preferences.

Here is the user's proposal: {proposal}. Based on this proposal, select one choice for each of the {num_questions} questions in the candidate feature checklist below.

{questions_text}

## Your strategic plan

You previously analyzed the design questions and produced the following plan. Use it to guide your questioning, but revise your approach as you learn from each user response.

{plan_text}

## User's Working Style (from prior projects)

The conversation summary below shows how this user approached PREVIOUS projects. Use it to understand their general preferences, values, and constraints.

{conversation_memory}

## Your task

You have at most {num_turns} rounds to ask the user multiple-choice questions via the `query_user' tool. You may ask any question you like - including questions taken directly from the checklist above (with their exact options), or your own original questions designed to learn about the user's preferences, values, and decision-making style. Your goal is to predict the user's answers to all {num_questions} questions above.

Use the `thought' field in `query_user' to record: (1) what you learned from the previous answer, (2) how it updates your understanding of the user, (3) whether and how you're revising your plan based on new information, and (4) why you're asking this next question.

The user may select multiple options, one option, or "None" followed by a brief explanation of what they'd prefer instead.

When you are confident enough to predict all {num_questions} questions in the checklist - or have used all {num_turns} query rounds - submit your spec sheet by calling `submit_predictions' with a prediction for every question (Q1 through Q{num_questions}). You MUST include all {num_questions} questions.

Do not use emoji, decorative formatting, or filler phrases (e.g., "Great question!", "Absolutely!"). Be direct and substantive.
\end{promptbox}

\vspace*{5mm}
We provide the system prompt for Task 2 planning mode below.
\vspace*{1mm}

\begin{promptbox}
You are an assistant helping a user produce a clear and structured specification sheet for their project. Before you begin the conversation, you need to plan your approach.

## Project Proposal
{proposal}

## Spec Template
The final specification must include these sections:
{sections_detail}

{memory_section}
## Your task
Analyze the proposal and produce a strategic questioning plan. You will have {num_turns} rounds to ask the user multiple-choice questions. Consider:

1. **What you already know** - What can you infer from the proposal alone?
2. **What is missing or ambiguous** - What key details are not addressed?
3. **Which spec sections need the most clarification** - Prioritize sections where the proposal gives the least guidance.
4. **Your questioning strategy** - Plan how to use your {num_turns} rounds efficiently. Each round you ask one multiple-choice question with exactly 4 options (A-D). Prioritize high-impact areas.

Output your plan as structured text. Keep it under 1000 words. Be concise:
- List only the key gaps and your questioning strategy
- Do not restate or summarize information already in the proposal
- Do not analyze sections that are already well-covered by the proposal

This is the planning phase only - you have no tools and no access to the user right now. Do not switch to execution mode, do not write as if you have already asked questions. Output only your plan.
\end{promptbox}

\vspace*{5mm}
We provide the system prompt for Task 2 execution mode below.
\vspace*{1mm}

\begin{promptbox}
You are an assistant helping a user produce a clear and structured specification sheet for their project. You have already analyzed the proposal and created a strategic plan.

## Project Proposal
{proposal}

## Spec Template
The final specification must include these sections:
{sections_detail}

{memory_section}
## Your Strategic Plan
{plan_text}

## Your task

You have at most {num_turns} rounds to ask the user multiple-choice questions via the `query_user' tool. You may ask any question you like - your own original questions designed to learn about the user's preferences, values, and decision-making style. Your goal is to gather enough information to produce a comprehensive specification covering all 9 sections.

Use the `thought' field in `query_user' to record: (1) what you learned from the previous answer, (2) how it updates your understanding of the user, (3) whether and how you're revising your plan based on new information, and (4) why you're asking this next question.

The user may select multiple options, one option, or "None" followed by a brief explanation of what they'd prefer instead.

When you have enough information - or at any time you feel confident - call the `submit_spec' tool to produce the final specification. When you have used all {num_turns} query rounds, call `submit_spec' to submit the final specification. All 9 sections are required.

## Guidelines
- Ask one question per round. Do not bundle multiple questions into a single message
- Each question must have exactly 4 options (option_a through option_d)
- Design options that are meaningfully distinct and cover the plausible design space
- Build on what the user says; don't repeat questions they've already answered
- Cover all spec sections; don't over-index on one area at the expense of others
- Follow your strategic plan, but adapt based on what you learn
- Do not use emoji, decorative formatting, or filler phrases (e.g., "Great question!", "Absolutely!"). Be direct and substantive
- Use ONLY the MCP tools (query_user, submit_spec). Do not use shell, file, or search tools
\end{promptbox}

\subsection{Evaluation}

We present the prompts used to evaluate the quality of the specification sheet produced by the agent for Task 2.

\vspace*{3mm}

\begin{promptbox}
You are role-playing as {persona[`name']}, a {persona[`age']}-year-old 
{persona[`occupation']} from {persona[`location']}.

## Your Background
{persona[`backstory']}

## Your Personality Traits
{traits_str}

## Your Behavioral Notes
{behavioral_notes}

## Your Project
{proposal}

## Your Requirements & Preferences
{requirements}

You will be given a specification sheet produced by an assistant for your project. Evaluate it from your perspective as {persona['name']}, considering your requirements and preferences above.

Here is the specification sheet produced by an assistant for your project:

{spec_text}
---

## How to evaluate
First, re-read your Requirements & Preferences above carefully. For each requirement you stated, check whether the spec addresses it. For each thing you said you DON'T want, check whether the spec avoids it.

Use this requirement-by-requirement and conversation review to ground your Coverage and Precision scores.

## Evaluation Criteria
Score the spec on these criteria (1-5 scale):

1. **Coverage** (1-5): How well the spec addresses each requirement and preference you stated above.
   - 1: The spec ignores most of your stated requirements
   - 2: A few requirements are addressed but major ones are missing
   - 3: Core requirements are present but several important preferences are missing
   - 4: Nearly all your requirements and preferences are reflected, with minor omissions
   - 5: Every requirement and preference you stated is clearly addressed in the spec

2. **Precision** (1-5): Whether the spec avoids including things you explicitly said you DON'T want, and avoids contradicting your stated preferences.
   - 1: Multiple items directly contradict your stated requirements or include things you explicitly rejected
   - 2: A few items contradict or include things you don't want
   - 3: Mostly aligned, but a few items feel off-base given your requirements
   - 4: Nearly everything aligns with your requirements; at most one minor misalignment
   - 5: Perfect alignment - nothing contradicts your requirements and nothing unwanted is included

3. **Internal Consistency** (1-5): Whether there are contradictions among items within the spec.
   - 1: Major contradictions between sections
   - 2: Several noticeable inconsistencies that undermine coherence
   - 3: Minor inconsistencies that could cause confusion
   - 4: Mostly coherent with at most one trivial inconsistency
   - 5: All sections are internally coherent with no contradictions

4. **Insight** (1-5): Whether the spec includes useful details, considerations, or recommendations that go beyond what you explicitly stated in your requirements.
   - 1: The spec only restates what you already said; no new useful details
   - 2: The spec adds one or two minor details, but nothing you hadn't already considered
   - 3: The spec includes a few useful additions you hadn't explicitly stated, such as edge cases, practical constraints, or sensible defaults
   - 4: The spec consistently adds valuable details - it surfaces considerations you would have eventually needed but hadn't thought to ask for
   - 5: The spec demonstrates deep understanding of your goals and includes specific, actionable recommendations you wouldn't have come up with yourself

5. **Readability** (1-5): How clear, concise, well-structured, and easy to understand the spec is.
   - 1: Disorganized, verbose, or difficult to follow
   - 2: Somewhat organized but unclear in places or unnecessarily wordy
   - 3: Reasonably clear but could be more concise or better structured
   - 4: Well-structured and clear with only minor areas for improvement
   - 5: Clear, concise, well-structured, and easy to act on

## Output Format
Respond with ONLY a JSON object in this exact format (no other text):
```json
{{
  "scores": {{
    "coverage": {{"score": <1-5>, "reason": "<1-2 sentences>"}},
    "precision": {{"score": <1-5>, "reason": "<1-2 sentences>"}},
    "consistency": {{"score": <1-5>, "reason": "<1-2 sentences>"}},
    "insight": {{"score": <1-5>, "reason": "<1-2 sentences>"}},
    "readability": {{"score": <1-5>, "reason": "<1-2 sentences>"}}
  }}
}}
'''

Be honest and specific. Think about what you would actually want as {persona_name}, given your background, requirements, and priorities.
\end{promptbox}

\section{\texttt{Buddy} Workflow Details}

We provide all details of the \texttt{Buddy} workflow, along with the prompts we wrote, for reproducing our results.

\subsection{User Profile}

We extract a user profile from the user's prior interactions with a coding agent. This profile is then used by user simulators to infer the user's preferences. An example of a user profile item is:
\begin{promptbox}
\key{"category"}: "priorities",
\key{"attribute"}: "cost sensitivity",
\key{"value"}: "Highly cost-conscious; tracks recurring subscription costs, compares monthly fees, and seeks to minimize or eliminate them where possible.",
\key{"type"}: "hard constraint",
\key{"confidence"}: 0.95,
\key{"evidence"}: [
{
  \key{"kind"}: "quote",
  \key{"text"}: "I've looked at stuff like Mindbody but honestly, those are way too expensive for what I need."
}
\end{promptbox}
\vspace*{3mm}
We define six dimensions of category which are useful attributes for inferring a user's preference for software development:
\begin{itemize}[leftmargin=1em]
\item working style: how the user thinks, communicates, scopes work, evaluates trade-offs, and makes progress,
\item priorities: what the user optimizes for,
\item design preferences: visual/UX/content-structure preferences that could generalize to other interfaces, 
\item technical fluency: the user's general technical fluency and comfort with complexity, 
\item domain background: high-level professional or domain background that may affect preferences,
\item decision rules: concrete heuristics the user appears to apply when making choices. 
\end{itemize}

Each item is assigned a type: a \emph{fact}, a \emph{soft preference}, or a \emph{hard constraint}. We also collect supporting evidence, which is either a direct quote from the user or a paraphrase of their statements from past conversations.

We provide the prompt used to extract a user's profile from their past conversation history below.
\vspace*{3mm}
\begin{promptbox}
You are an expert user researcher analyzing conversation transcripts. You are extracting a reusable user preference profile from a conversation between a user and a coding agent.

Your goal is NOT to summarize the project.
Your goal is to infer the user's likely preferences for FUTURE software/product/design decisions.

## Conversation
{conv_text}

## What to extract
Extract only attributes that are likely to transfer to a different project, product, or design context.

Prioritize:
- stable behavioral patterns
- recurring decision rules
- durable values and priorities
- reusable UX / aesthetic preferences
- general technical sophistication
- high-level domain background that may shape preferences

Do NOT prioritize:
- project-specific implementation details
- one-off requests
- accidental constraints caused by the current task
- details that would not help predict choices on a new project

## Core rules
1. Evidence first.
   Every attribute must be grounded in evidence from the conversation.

2. Be conservative.
   If evidence is weak, conflicting, or too project-specific, omit the attribute.

3. Prefer repeated or choice-revealing signals.
   Strongest signals come from:
   - explicit stated preferences
   - repeated requests across the conversation
   - trade-offs the user chose
   - things the user rejected
   - corrections the user made to the agent

4. Separate facts from preferences.
   - "fact" = something true about the user/context
   - "soft_preference" = likely tendency, but not absolute
   - "hard_constraint" = explicit must-have / must-not-have

5. Optimize for transferability.
   Ask: "Would this help predict what this user wants in a different software project?"
   If no, omit it.

6. Do not force coverage.
   It is better to return fewer high-confidence attributes than many weak ones.

7. Avoid duplication.
   Merge overlapping attributes into one stronger attribute.

## Category definitions
Use only these categories:

1. working_style
   How the user thinks, communicates, scopes work, evaluates trade-offs, and makes progress.
   Examples: iterative vs upfront planning, detail orientation, pragmatism, decision speed, desire for explanation.

2. priorities
   What the user optimizes for.
   Examples: simplicity, speed, quality, polish, maintainability, cost sensitivity, reliability, usability, control.

3. design_preferences
   Visual/UX/content-structure preferences that could generalize to other interfaces.
   Examples: minimal vs dense, formal vs casual, clean layouts, low animation, mobile-first, information hierarchy.

4. technical_fluency
   The user's general technical fluency and comfort with complexity.
   Keep this general and transferable.
   Examples: highly technical, comfortable reviewing architecture, wants abstractions explained, prefers standard patterns.

5. domain_background
   High-level professional or domain background that may affect preferences.
   Keep this broad.
   Examples: healthcare professional, enterprise buyer mindset, consumer app operator.

6. decision_rules
   Concrete heuristics the user appears to apply when making choices.
   Each item should be phrased as a concrete rule that predicts future choices.
   Good format:
   - "Subtractive scope - defaults to removing non-core features unless they have clear immediate value."
   - "Standard-over-clever - prefers conventional, explainable solutions over abstract flexibility."
   - "Polish where visible - accepts technical shortcuts more than UX roughness."
   These must be inferred from actual decisions, accept/reject patterns, or trade-off discussions in the conversation.

## Attribute quality bar
Only include an attribute if at least one of these is true:
- the user explicitly stated it
- the user repeated it
- the user made a choice that clearly reveals it
- the user rejected an alternative in a way that reveals a general rule

Prefer attributes supported by 2+ pieces of evidence.
A single strong explicit statement is acceptable.

## Output requirements
Return valid JSON only. Use this exact schema:

{
  "attributes": [
    {
      "category": "working_style | priorities | design_preferences | technical_fluency | domain_background | decision_rules",
      "attribute": "snake_case_attribute_name",
      "value": "clear natural-language description of the inferred trait/rule",
      "type": "hard_constraint | soft_preference | fact",
      "confidence": "0.0 to 1.0 (numeric)",
      "evidence": [
        {
          "kind": "quote | paraphrase",
          "text": "supporting evidence from the conversation"
        }
      ],
      "reasoning": "1 short sentence explaining why this evidence supports a reusable preference rather than a one-off project detail"
    }
  ]
}

## Additional formatting rules
- Use snake_case for "attribute".
- "value" must be specific enough to predict future choices.
- Keep each "reasoning" sentence short and concrete.
- Do not invent evidence.
- Do not output empty placeholder attributes.
- Do not include project-specific technologies, libraries, APIs, schemas, compliance rules, or implementation details unless they reveal a broader transferable preference.
\end{promptbox}

\subsection{Draft A Spec Sheet}

The first step of \texttt{Buddy} is to generate a spec sheet draft from the proposal (i.e., the user's intent). To better align the draft with the user's preferences, it incorporates the extracted user profile and conversation memory. In addition to the spec sheet draft, \texttt{Buddy} also produces an assumption log, a list of open questions, and a project-specific multiple-choice question. We provide the prompt below.
\vspace*{3mm}
\begin{promptbox}
You are a senior product manager drafting a first-pass software specification from an early project proposal. Your job is to produce a useful draft that is clear about three things: (1) what the proposal explicitly states, (2) what you had to infer to make the spec coherent, and (3) which important decisions remain open. Do not invent unnecessary detail. Infer only what is needed. When a user profile is available, use it to make better-informed inferences about the user's likely preferences - but always defer to the proposal.

## Project Proposal
{proposal}

## User Profile (inferred from prior interactions)
{user_profile}

## User's General Preferences (from prior projects)
This summary shows how the user approached a PREVIOUS, UNRELATED project.
Use it to understand their general values, constraints, and working style.
The CURRENT project is defined by the Proposal above, not this summary.

{conversation_memory}

## Spec Sections
{sections_desc}

## Task
Produce a first-pass specification for this project.

For each of the 9 sections:
- write a concise but substantive draft based on the proposal
- incorporate only the minimum necessary inferred detail needed to make the section coherent
- when the user profile suggests a preference direction, lean toward it
- do not pad the section with generic best practices

Then identify:
1. **Assumptions**: consequential decisions or details that were not explicitly stated in the proposal but were inferred in order to write the spec
2. **Open questions**: important design decisions that should remain unresolved because multiple plausible choices exist

## What counts as a good assumption
A good assumption is:
- not explicitly stated in the proposal
- consequential to product scope, UX, architecture, operations, or delivery
- specific enough to validate with a stakeholder
- something a reasonable team might choose differently

Do NOT include trivial assumptions, generic software platitudes, or restatements of the proposal.

## Output format
Respond with valid JSON only:
{
  "spec": {
    "project_overview": "...",
    ... (all 9 sections)
  },
  "assumptions": [
    {
      "id": "A01",
      "text": "Assumed the product will launch first as a responsive web app.",
      "sections": ["project_overview", "user_experience", "technical_decisions"],
      "importance": "high"
    }
  ],
  "open_questions": [
    {
      "id": "Q01",
      "text": "Whether the initial release should support SSO or only email/password authentication.",
      "sections": ["technical_decisions", "user_experience"],
      "importance": "high"
    }
  ],
  "priority_mcq": {
    "question": "What matters most for the first version of this project?",
    "options": {
      "A": "Option A - a project-specific strategic direction",
      "B": "Option B - a different project-specific strategic direction",
      "C": "Option C - a third project-specific strategic direction",
      "D": "Option D - a fourth project-specific strategic direction"
    }
  }
}

## Writing guidelines
- Each section should be 2-5 sentences of clear, concise writing
- Prefer concrete, testable language
- Preserve ambiguity where the proposal genuinely leaves room for multiple legitimate choices
- Infer defaults only when necessary to avoid an unusably vague spec
- Include roughly 6-15 total assumptions, not an assumption list for every section
- Include roughly 4-10 open questions, focused on high-leverage design decisions
- Include exactly 1 priority MCQ with 4 project-specific options

The priority MCQ should ask: "What matters most for the first version of this project?" with 4 options that are SPECIFIC to this project (not generic). Each option should represent a different strategic direction that would meaningfully change the spec. Example for a watch party app: A) Sub-second sync accuracy, B) Broad device support, C) Rich social features from day one, D) Minimal viable product shipped in 4 weeks.
\end{promptbox}

\subsection{Morphological Decomposition}

We apply morphological analysis to decompose the spec sheet into several dimensions. Each dimension contains multiple options that users may prefer differently and therefore require user clarification. We provide the prompt below.

\vspace*{3mm}
\begin{promptbox}
You are a product analyst decomposing a project's requirement space. You identify the smallest set of high-value preference axes where different users would genuinely want different things - in terms of features, scope, user experience, and constraints - NOT implementation technology.

## Project Proposal
{proposal}

## User Profile
{user_profile}

## User's General Preferences (from prior projects)
{conversation_memory}

## Draft Specification
{spec_block - all 9 sections}

## Assumptions Made in Draft
{assumptions list}

## Open Questions
{open_questions list}

## Task
Decompose this project's requirement space into preference axes - dimensions where different users would genuinely want different things.

A **requirement dimension** is a preference axis where:
- multiple legitimate user preferences exist
- different users with different values, goals, or constraints would choose differently
- the choice materially affects what the product DOES, how it FEELS, or what it EXCLUDES

CRITICAL: Dimensions must describe WHAT the user wants (features, scope, behavior, UX preferences, constraints, boundaries), NOT HOW to implement it. Do NOT create dimensions about technology choices, frameworks, libraries, hosting providers, databases, or implementation architecture.

For each dimension:
- give it a concise name describing the user preference (e.g., "notification_preferences", not "frontend_framework")
- describe the preference it captures
- write a clear, concise user-facing question (under 50 words) a non-technical user would understand
- identify the approach the draft spec currently leans toward
- provide exactly 4 alternatives describing meaningfully different approaches to scope, behavior, or user experience
- note other dimensions it materially interacts with

## Prioritize dimensions that are:
- high leverage for the user's experience
- discriminative (different personas would genuinely pick different options)
- about features, scope boundaries, or UX preferences implied by assumptions or open questions
- useful for understanding what the user values vs what they'd cut

## Include 2-3 dimensions about EXCLUSIONS and SCOPE BOUNDARIES
Different users draw scope boundaries differently. Include dimensions like:
- What should be explicitly OUT of scope for the initial version?
- Which capabilities should be deferred vs included now?
- What complexity level is appropriate for the target user?

## Exclude dimensions that are:
- already explicitly fixed by the proposal
- about implementation technology (frameworks, languages, hosting, databases, APIs, libraries)
- too broad to guide decisions
- too narrow or cosmetic
- simple sub-choices of a larger dimension already captured elsewhere
- about strategic or business meta-decisions, not user-facing features. Specifically, do NOT create dimensions about:
  - MVP scope boundaries or release phasing ("what ships first vs later")
  - Pricing tiers, paywall placement, or freemium structure
  - Budget, cost, or resource constraints
  - Build timeline or onboarding duration
  - "Complexity ceiling" or "how much to build"
  These are strategic decisions that should be handled as constraints, not as preference dimensions. Only create dimensions about what the product does and how the user experiences it.

## Scope constraint
Every dimension MUST be grounded in something the proposal explicitly states or clearly implies. If the proposal does not mention a capability, do not create a dimension about it.

## Target size
Return the **best 8-12 dimensions total** for the whole project, not a quota per section.
Focus on the highest-leverage dimensions where different users would genuinely disagree.
**Order dimensions by importance to the user** (most important first).

## Output JSON
Respond with valid JSON only:
[
  {
    "dimension_key": "notification_preferences",
    "primary_section": "core_features",
    "related_sections": ["user_experience", "non_functional_requirements"],
    "question": "How should the system notify users about important updates?",
    "description": "How much control users have over when and how they receive notifications.",
    "assumed_approach": "Email notifications for critical updates only",
    "why_it_matters": "Too many notifications annoy users; too few cause them to miss important information.",
    "alternatives": [
      "No notifications - users check the system manually when they want",
      "Email notifications for critical updates only (e.g., status changes)",
      "Email and in-app notifications with user-configurable frequency",
      "Real-time push notifications for all activity with per-category controls"
    ],
    "depends_on": ["user_engagement_level", "mobile_experience_priority"]
  }
]

## Quality bar
- Prefer fewer, better dimensions over exhaustive coverage
- Every alternative must be understandable by a non-technical user
- Ground dimensions in the proposal, spec, assumptions, and open questions
- Test each dimension: "Would a non-technical end user have a meaningful opinion on this?" If not, it's too technical - reframe or drop it.

## Overlap prevention (critical)
Before finalizing, review your dimensions for semantic overlap. Two dimensions overlap if a user's answer to one would largely determine their answer to the other. Common traps:
- Same topic at different granularities (e.g., where something lives vs how it is managed)
- Same decision reframed with different wording
- Implementation detail that follows directly from a broader architecture choice

If you find overlap, keep the broader/higher-leverage dimension and drop the narrower one. Every dimension in your output must be independently answerable without duplicating another dimension's decision.
\end{promptbox}

\subsection{Simulated User Preliminary Decisions}

We create three simulated user agents to filter out morphological dimensions that can be easily inferred from a user's preferences based on their past conversation history. Each simulated agent is assigned a different system prompt to encourage diverse reasoning behaviors. For every morphological dimension, the agents predict which option the user is most likely to prefer. During the user interview, we prioritize questions (i.e., morphological dimensions) for which the agents either disagree on the predicted choice or exhibit low confidence in their predictions.
We provide different system prompts for three simulated users. 
\vspace*{3mm}
\begin{promptbox}
### System Prompt Variant 0 (full user profile)

You are predicting a user's requirement preferences across multiple dimensions of a software project. For each preference axis, select what this user would most likely want given their profile, project context, and behavioral patterns.

### System Prompt Variant 1 (20

Role-play as the user described below. For each preference axis, choose what you would naturally want. Think about how someone with these characteristics, goals, and constraints would feel about each feature scope and UX decision.

### System Prompt Variant 2 (profile-free)

Based on this user's profile and project context, determine their preferred scope and experience for each axis. Consider their values, comfort level, budget, and stated priorities when predicting each preference.
\end{promptbox}

\vspace*{3mm}
We provide the user prompt shared by all simulated users below. 

\vspace*{3mm}

\begin{promptbox}
## Project Proposal
{proposal}

## Person's Background & Behavioral Patterns (from prior interaction)
{conversation_memory}

## User Profile (inferred - may be incomplete)
{user_profile}

## Design Dimensions
For each dimension below, predict which alternative approach this user would prefer.

{dim_text - compact dimension listing with alternatives}

## Instructions
For each dimension, THINK FIRST before answering:
1. Consider the user's profile, project proposal, and behavioral patterns
2. Evaluate each alternative against what you know about this user
3. Select the single approach they would most likely prefer

Assess your confidence: "confident" is true if the profile gives enough signal for a reliable prediction, false if you're guessing due to insufficient information.

Respond with a JSON array. Use the exact dimension ID strings shown above. Put "reasoning" BEFORE "choice":
[{"dimension": "notification_preferences", "reasoning": "This user values simplicity and has limited technical skill, so the spreadsheet-driven approach fits best.", "choice": "B", "confident": true}]

Reasoning should be 1-2 sentences referencing concrete profile attributes, goals, or stated preferences. Include an entry for every dimension listed above.
\end{promptbox}

\subsection{Writing and Reviewing the Final Spec Sheet}

Finally, after interviewing the user, the agent drafts the final spec sheet using the prompt provided below.

\vspace*{3mm}
\begin{promptbox}
You are a senior product manager writing the final specification for a software project. You produce clear, actionable, well-structured specs that are ready for a development team to execute.

## Project Proposal
{proposal}

## User Profile
{profile_attrs as JSON}

## User Corrections
The user clarified {len(novel_info_all)} preferences during the conversation:
{corrections_list}

These preferences take priority over inferred and consensus decisions below.

## User Priorities and Constraints
Q: {priority question}
A: {priority answer}

## Conversation with User
{interaction_text - one line per turn summarizing dimension + choice}

## Resolved Design Decisions
These are the confirmed design decisions, organized by section. "USER CONFIRMED" means the user explicitly stated this preference. "high confidence" means the profile strongly indicates this. "inferred" means this is a best guess.

{decisions_text - grouped by spec section with priority tags, rejected alternatives, and why-it-matters context}

## User Corrections and Custom Preferences
The user explicitly stated these preferences. They override any inferred decisions:
{corrections_display}

## Task
Write the final specification. Produce all 9 sections as clear, actionable prose.

## SCOPE CONTROL - DO NOT INCLUDE UNLESS EXPLICITLY REQUESTED
The following categories are common sources of scope creep. Do NOT include ANY unless the proposal explicitly describes them or the user explicitly requested them:
- Admin panels, dashboards, or management interfaces
- Analytics, reporting, metrics tracking, or usage statistics
- Notification systems (email, push, in-app, SMS)
- Social features, sharing, collaboration, or community
- Gamification, achievements, badges, or rewards
- Multiple user roles beyond what the proposal states
- API endpoints, webhooks, or third-party integrations not mentioned
- Advanced settings, customization panels, or preference pages
- Audit logging, compliance tracking, or activity feeds
- Multi-language/i18n support (unless mentioned)
- Mobile apps (unless the proposal specifically mentions mobile)

For each feature in your spec, check: "Did the proposal or user ask for this?" If no, remove it.

Guidelines:
- USER CONFIRMED decisions are absolute - never contradict them
- High confidence decisions should be stated as definitive
- Inferred decisions: Include ONLY if the choice is an obvious, non-controversial default for this type of project. If the inference is genuinely uncertain or could easily go either way, OMIT the decision from the spec entirely rather than guessing wrong - a gap is better than a wrong commitment. When you do include an inferred decision, state it as a recommended default (e.g., "initially targets...", "starts with...") without labeling it as an assumption.
- Use "Why it matters" context to ground decisions (a brief phrase, not a full explanation)
- Use "Rejected alternatives" to briefly note key trade-offs where relevant
- Multi-selection: If a dimension has multiple selected options (e.g., "A + B"), the user indicated interest in both approaches. Where compatible, describe a hybrid or combined approach. Where they conflict, pick the approach that better fits the user's other confirmed preferences and note the trade-off.
- Phase coherence rule: Before writing milestones_and_priorities, establish a single canonical list of what is "core/Phase 1" vs "deferred/later". Every other section must be consistent with this list.
- out_of_scope rules: Only list genuinely excluded CAPABILITIES - things a stakeholder might reasonably expect but that are NOT being built.
- Section cross-check: After drafting all sections, verify: (a) every item in core_features appears in milestones, (b) no item in out_of_scope appears as a deliverable elsewhere, (c) constraints mentioned in one section are not contradicted in another.
- Coverage gap check: Every design decision marked [USER CONFIRMED] in the Resolved Design Decisions MUST be reflected in the relevant spec section.
- Proposal alignment: For features NOT mentioned in user confirmed requirements, check alignment with the proposal. If the proposal does NOT mention a capability and the user did not explicitly request it, do not include it. Do not relabel removed items as "Phase 2" or "optional" - remove them entirely.
- Each section should be 2-5 sentences; prefer bullet points for lists
- Be concise. State decisions and requirements directly. Do not include extended rationale or justification paragraphs.
- HARD LIMIT: The total spec MUST be under 6,000 characters across all 9 sections.
- Language calibration: If the user's profile indicates limited technical background, never name specific frameworks, libraries, algorithms, or infrastructure tools. Describe capabilities in terms of what the user experiences, not how the system implements it.
- No internal markup in output: Do NOT include source tags like [USER CONFIRMED], [high confidence], or [inferred] in your output text.
- Practical grounding: Include 2-3 specific, practical implementation details or domain recommendations the user didn't explicitly request but would strengthen the spec.
- Proposal feature floor: The proposal defines the minimum feature set. User responses during conversation shape HOW features work but should not be used to remove features the proposal explicitly describes.

Respond with valid JSON only:
{
  "project_overview": "...",
  ... (all 9 sections)
}
\end{promptbox}

\vspace*{3mm}
The final step is to review the spec sheet with the prompt given below.
\vspace*{3mm}
\begin{promptbox}
You are a senior product manager reviewing a draft specification. Your job is to catch errors and produce a corrected version.

## Project Proposal
{proposal}

## User Profile
{profile_attrs as JSON}

## User Interaction Summary
{interaction_summary}

## User Corrections
{corrections list}

## Draft Specification
{spec as JSON}

## Review Checklist
Check the draft specification for these issues and fix any you find:

1. **Coverage**: Every feature or capability mentioned in the proposal should appear somewhere in the spec. List any proposal features missing from the spec and add them.

2. **Precision / scope creep**: Every feature in the spec should trace back to the proposal or a user interaction. Remove features, roles, or workflows that were never discussed (e.g., admin panels, analytics dashboards, notification systems, account registration not mentioned in the proposal).

3. **Explicit requirement preservation**: If the user or proposal called something "core", "must-have", or "do not cut", verify it appears in Phase 1 / MVP milestones - not deferred to later phases.

4. **Consistency**: Check for internal contradictions - features included in one section but excluded in another, conflicting technical decisions, milestones that don't match core_features.

5. **Operational specificity**: Where performance targets exist, ensure they use percentile notation (p95/p99). For critical components, include a brief failure mode description. Ensure milestones are realistic for the stated team size.

Output the corrected specification as valid JSON with the same 9 section keys. Only change sections that have actual issues - do not rewrite sections that are fine. Keep the same concise style. Stay within 6,000 characters total.
\end{promptbox}

\subsection{An Example Spec Sheet}

We show an example spec sheet \texttt{Buddy} wrote for the ml-fairness (bias auditing tool) project.

\begin{tcolorbox}[title=Project Overview]
A classification-focused ML fairness and bias auditing tool that computes standard fairness metrics across protected attributes, supports intersectional analysis, and presents results through a polished interactive web application.

The tool wraps established libraries (Fairlearn/AIF360) behind a unified workflow, adding value through visualization, what-if mitigation controls, model card generation, and a built-in demo experience using the Adult Income dataset.

The primary goal is a compelling, recruiter-ready portfolio piece optimized for a 10-minute live walkthrough demonstrating deep understanding of ML fairness concepts.
\end{tcolorbox}

\begin{tcolorbox}[title=Core Features]
\begin{itemize}[leftmargin=0em]
    \item \textbf{Fairness Metric Computation}: Comprehensive classification metrics (demographic parity, equalized odds, predictive parity, calibration, disparate impact ratio, and others) computed across user-specified protected attributes. Supports intersectional analysis for up to three-way group combinations.

    \item \textbf{Interactive Dashboard}: Full web application with a high-level summary view that lets users drill down into per-group and intersectional breakdowns. Contextual tooltips provide brief metric explanations on hover without cluttering the interface.

    \item \textbf{What-If Mitigation Controls}: One post-processing mitigation technique (threshold optimization via Fairlearn) with interactive sliders showing fairness-accuracy trade-off curves in real time. Users adjust constraints and immediately see updated metric values.

    \item \textbf{Black-Box Auditing}: The tool operates from predictions and labels alone by default. No white-box features (e.g., SHAP) in scope.

    \item \textbf{Model Card Export}: Auto-generate model card sections (fairness metrics, dataset demographics, known limitations) from audit results, exportable as Markdown.

    \item \textbf{Built-In Demo}: Preloaded Adult Income dataset with a trained classifier and a ``Try It Now'' button that runs a full audit instantly. This is the default landing experience.

    \item \textbf{Bias Drift Detection}: Monitoring module that accepts incoming prediction batches, computes fairness metrics over time windows, and raises alerts when metrics drift beyond configured thresholds. Detection and alerting only --- no automated remediation.
\end{itemize}
\end{tcolorbox}

\begin{tcolorbox}[title=Non-Functional Requirements]
\begin{itemize}[leftmargin=0em]
    \item The built-in demo audit (Adult Income, \textasciitilde32K rows) must complete and render results within 5 seconds (p95) to maintain demo flow.

    \item The frontend must render smoothly at 60fps during slider interactions on the what-if trade-off curves.

    \item What-if slider adjustments must reflect updated metrics within 200ms (p95).

    \item All free-tier infrastructure; no paid services required to run locally or deploy a demo instance.

    \item Failure modes:
    \begin{itemize}[leftmargin=0em]
        \item If the metric engine errors on a group (e.g., division by zero for empty subgroups), surface a clear warning per-group rather than failing the entire audit.
        \item If bias drift ingestion fails, queue the batch and retry with an alert after 3 consecutive failures.
    \end{itemize}
\end{itemize}
\end{tcolorbox}

\begin{tcolorbox}[title=Technical Decisions]
\begin{itemize}[leftmargin=0em]
    \item \textbf{Algorithm Layer}: Thin integration over Fairlearn and AIF360 --- the tool's value is in workflow, visualization, and UX, not reimplemented algorithms. Wrap library calls behind a unified internal API so the frontend has a single consistent interface.

    \item \textbf{Mitigation Technique}: Fairlearn's ThresholdOptimizer for post-processing --- it produces interpretable threshold adjustments per group and pairs naturally with interactive trade-off visualization.

    \item \textbf{Architecture}: Python FastAPI backend serving a React frontend. The backend exposes a REST API for audit runs, metric queries, and mitigation simulations. The frontend is a standalone SPA.

    \item \textbf{Demo Dataset}: Adult Income (UCI) with a pre-trained gradient boosted classifier shipped as a static asset, so the demo runs without any model training.

    \item \textbf{Intersectional Groups}: Cap group combinations at three attributes. Warn users when any intersectional subgroup has fewer than 30 samples, as metric estimates become unreliable below this threshold.
\end{itemize}
\end{tcolorbox}

\begin{tcolorbox}[title=User Experience]
\begin{itemize}[leftmargin=0em]
    \item \textbf{Landing Page}: Clean hero section with project description and a prominent ``Try It Now'' button that loads the pre-built Adult Income audit.

    \item \textbf{Dashboard Layout}: Three-tier drill-down:
    \begin{enumerate}
        \item Summary scorecard with pass/fail indicators per metric
        \item Per-group bar charts for a selected metric
        \item Intersectional heatmap for multi-attribute breakdowns
    \end{enumerate}

    \item \textbf{Tooltips}: Every metric label has a hover tooltip with a one-sentence plain-language explanation and the mathematical definition. Tooltips stay out of the way until invoked.

    \item \textbf{What-If Panel}: Side panel with constraint sliders (e.g., maximum demographic parity difference). Adjusting a slider updates a Pareto-style accuracy-vs-fairness curve and highlights the current operating point.

    \item \textbf{Model Card Tab}: Read-only preview of the auto-generated model card with a ``Download as Markdown'' button.

    \item \textbf{Upload Flow}: For custom audits --- upload a CSV of predictions/labels/attributes. Validate schema on upload with clear error messages.
\end{itemize}
\end{tcolorbox}

\begin{tcolorbox}[title=Out of Scope]
\begin{itemize}[leftmargin=0em]
    \item Regression and ranking model fairness analysis
    \item Causal fairness analysis (DAGs, counterfactual fairness, path-specific effects)
    \item Pre-processing and in-processing mitigation techniques
    \item Automated bias remediation (auto-applying fixes to production models)
    \item Production deployment integration (CI/CD hooks, model registry connectors)
    \item Multi-user accounts or authentication
    \item Persistent storage of historical audits across sessions
    \item SHAP or other feature attribution analysis
    \item White-box model internals beyond what black-box predictions provide
\end{itemize}
\end{tcolorbox}

\begin{tcolorbox}[title=Milestones and Priorities]

\textbf{Phase 1 --- Metric Engine + API}\\
\textit{Tests written alongside each module}

\begin{itemize}[leftmargin=0em]
    \item Fairlearn/AIF360 integration layer: compute all classification fairness metrics for single and intersectional groups
    \item FastAPI endpoints for running an audit, querying metrics, and fetching group breakdowns
    \item Unit tests validating metric correctness against direct library calls
\end{itemize}

\vspace{0.5em}

\textbf{Phase 2 --- Dashboard Core}

\begin{itemize}[leftmargin=0em]
    \item React frontend: summary scorecard, per-group bar charts, intersectional heatmap
    \item Contextual tooltip system
    \item CSV upload flow with schema validation
    \item Wire to backend API; end-to-end test with Adult Income data
\end{itemize}

\vspace{0.5em}

\textbf{Phase 3 --- What-If Mitigation + Demo}

\begin{itemize}[leftmargin=0em]
    \item ThresholdOptimizer integration with API endpoint returning trade-off curve data
    \item Interactive slider panel with real-time chart updates
    \item Pre-built Adult Income demo with ``Try It Now'' button
    \item Frontend black-box-only mode (no model object required)
\end{itemize}

\vspace{0.5em}

\textbf{Phase 4 --- Model Card, Monitoring, Polish}

\begin{itemize}[leftmargin=0em]
    \item Model card auto-generation and Markdown export
    \item Bias drift detection module: accepts prediction batches, computes windowed metrics, surfaces threshold-based alerts (detection and alerting only)
    \item README, deployment to free-tier hosting (Render), final UI polish
\end{itemize}

\end{tcolorbox}

\begin{tcolorbox}[title=Success Metrics]
\begin{itemize}[leftmargin=0em]
    \item A cold visitor can complete the built-in demo audit and view all dashboard tabs within 2 minutes of landing on the page.

    \item The full 10-minute walkthrough covers:
    \begin{itemize}[leftmargin=1em]
        \item Demo audit
        \item Metric drill-down
        \item Intersectional heatmap
        \item What-if mitigation
        \item Model card export
    \end{itemize}
    without any loading delays or dead ends.

    \item The README and live demo are sufficient for a technical interviewer to understand the project's scope, architecture, and fairness domain knowledge without running the code locally.

    \item All fairness metric values match Fairlearn's direct output on the same data (validated by automated tests).
\end{itemize}
\end{tcolorbox}

\begin{tcolorbox}[title=Constraints]
\begin{itemize}[leftmargin=0em]
    \item Zero ongoing cost: all infrastructure on free tiers (Render, SQLite for any persistence). No paid APIs.

    \item Solo developer, intermediate Python skill level --- scope is calibrated to ship a polished, complete project rather than attempt breadth.

    \item Tests are written alongside each feature module, not deferred to a separate phase.

    \item The tool must function fully in black-box mode (predictions + labels only). No white-box model access features in v1.
\end{itemize}
\end{tcolorbox}

%% file: main.bbl
\newcommand{\etalchar}[1]{$^{#1}$}
\begin{thebibliography}{WWY{\etalchar{+}}24}

\bibitem[AAA{\etalchar{+}}23]{achiam2023gpt}
Josh Achiam, Steven Adler, Sandhini Agarwal, Lama Ahmad, Ilge Akkaya,
  Florencia~Leoni Aleman, Diogo Almeida, Janko Altenschmidt, Sam Altman,
  Shyamal Anadkat, et~al.
\newblock Gpt-4 technical report.
\newblock {\em arXiv preprint arXiv:2303.08774}, 2023.

\bibitem[AFGG24]{andukuri2024star}
Chinmaya Andukuri, Jan-Philipp Fr{\"a}nken, Tobias Gerstenberg, and Noah~D
  Goodman.
\newblock Star-gate: Teaching language models to ask clarifying questions.
\newblock {\em arXiv preprint arXiv:2403.19154}, 2024.

\bibitem[CAL94]{cohn1994improving}
David Cohn, Les Atlas, and Richard Ladner.
\newblock Improving generalization with active learning.
\newblock {\em Machine learning}, 15(2):201--221, 1994.

\bibitem[CKA{\etalchar{+}}25]{chhikara2025mem0}
Prateek Chhikara, Dev Khant, Saket Aryan, Taranjeet Singh, and Deshraj Yadav.
\newblock Mem0: Building production-ready ai agents with scalable long-term
  memory.
\newblock {\em arXiv preprint arXiv:2504.19413}, 2025.

\bibitem[CXL{\etalchar{+}}23]{chuang2023dola}
Yung-Sung Chuang, Yujia Xie, Hongyin Luo, Yoon Kim, James Glass, and Pengcheng
  He.
\newblock Dola: Decoding by contrasting layers improves factuality in large
  language models.
\newblock {\em arXiv preprint arXiv:2309.03883}, 2023.

\bibitem[DHF{\etalchar{+}}25]{deng2025interactcomp}
Mingyi Deng, Lijun Huang, Yani Fan, Jiayi Zhang, Fashen Ren, Jinyi Bai, Fuzhen
  Yang, Dayi Miao, Zhaoyang Yu, Yifan Wu, et~al.
\newblock Interactcomp: Evaluating search agents with ambiguous queries.
\newblock {\em arXiv preprint arXiv:2510.24668}, 2025.

\bibitem[DPS{\etalchar{+}}25]{damani2025beyond}
Mehul Damani, Isha Puri, Stewart Slocum, Idan Shenfeld, Leshem Choshen, Yoon
  Kim, and Jacob Andreas.
\newblock Beyond binary rewards: Training lms to reason about their
  uncertainty.
\newblock {\em arXiv preprint arXiv:2507.16806}, 2025.

\bibitem[ES26]{edwards2026ask}
Nicholas Edwards and Sebastian Schuster.
\newblock Ask or assume? uncertainty-aware clarification-seeking in coding
  agents.
\newblock {\em arXiv preprint arXiv:2603.26233}, 2026.

\bibitem[FCQ{\etalchar{+}}24]{feng2024large}
Xueyang Feng, Zhi-Yuan Chen, Yujia Qin, Yankai Lin, Xu~Chen, Zhiyuan Liu, and
  Ji-Rong Wen.
\newblock Large language model-based human-agent collaboration for complex task
  solving.
\newblock In {\em Findings of the Association for Computational Linguistics:
  EMNLP 2024}, pages 1336--1357, 2024.

\bibitem[FLC17]{fang2017learning}
Meng Fang, Yuan Li, and Trevor Cohn.
\newblock Learning how to active learn: A deep reinforcement learning approach.
\newblock In {\em Proceedings of the 2017 conference on empirical methods in
  natural language processing}, pages 595--605, 2017.

\bibitem[GG16]{gal2016dropout}
Yarin Gal and Zoubin Ghahramani.
\newblock Dropout as a bayesian approximation: Representing model uncertainty
  in deep learning.
\newblock In {\em international conference on machine learning}, pages
  1050--1059. PMLR, 2016.

\bibitem[GLW{\etalchar{+}}24]{gao2024enhancing}
Yiming Gao, Feiyu Liu, Liang Wang, Zhenjie Lian, Dehua Zheng, Weixuan Wang,
  Wenjin Yang, Siqin Li, Xianliang Wang, Wenhui Chen, et~al.
\newblock Enhancing human experience in human-agent collaboration: a
  human-centered modeling approach based on positive human gain.
\newblock {\em arXiv preprint arXiv:2401.16444}, 2024.

\bibitem[HWM25]{hu2025evaluating}
Yuanzhe Hu, Yu~Wang, and Julian McAuley.
\newblock Evaluating memory in llm agents via incremental multi-turn
  interactions.
\newblock {\em arXiv preprint arXiv:2507.05257}, 2025.

\bibitem[JYL{\etalchar{+}}24]{jia2024can}
Feiran Jia, Ziyu Ye, Shiyang Lai, Kai Shu, Jindong Gu, Adel Bibi, Ziniu Hu,
  David Jurgens, James Evans, Philip~H Torr, et~al.
\newblock Can large language model agents simulate human trust behavior?
\newblock {\em Advances in neural information processing systems},
  37:15674--15729, 2024.

\bibitem[KICB25]{kirichenko2025abstentionbench}
Polina Kirichenko, Mark Ibrahim, Kamalika Chaudhuri, and Samuel~J Bell.
\newblock Abstentionbench: Reasoning llms fail on unanswerable questions.
\newblock {\em arXiv preprint arXiv:2506.09038}, 2025.

\bibitem[KNVZ25]{kalai2025language}
Adam~Tauman Kalai, Ofir Nachum, Santosh~S Vempala, and Edwin Zhang.
\newblock Why language models hallucinate.
\newblock {\em arXiv preprint arXiv:2509.04664}, 2025.

\bibitem[LHE22]{lin2022teaching}
Stephanie Lin, Jacob Hilton, and Owain Evans.
\newblock Teaching models to express their uncertainty in words.
\newblock {\em arXiv preprint arXiv:2205.14334}, 2022.

\bibitem[LHL{\etalchar{+}}23]{lu2023does}
Bo-Ru Lu, Nikita Haduong, Chia-Hsuan Lee, Zeqiu Wu, Hao Cheng, Paul Koester,
  Jean Utke, Tao Yu, Noah~A Smith, and Mari Ostendorf.
\newblock Does collaborative human-lm dialogue generation help information
  extraction from human dialogues?
\newblock {\em arXiv preprint arXiv:2307.07047}, 2023.

\bibitem[LTGA23]{li2023eliciting}
Belinda~Z Li, Alex Tamkin, Noah Goodman, and Jacob Andreas.
\newblock Eliciting human preferences with language models.
\newblock {\em arXiv preprint arXiv:2310.11589}, 2023.

\bibitem[LXY{\etalchar{+}}25]{li2025far}
Rui Li, Heming Xia, Xinfeng Yuan, Qingxiu Dong, Lei Sha, Wenjie Li, and Zhifang
  Sui.
\newblock How far are llms from being our digital twins? a benchmark for
  persona-based behavior chain simulation.
\newblock In {\em Findings of the Association for Computational Linguistics:
  ACL 2025}, pages 15738--15763, 2025.

\bibitem[MDCS25]{mysore2025prototypical}
Sheshera Mysore, Debarati Das, Hancheng Cao, and Bahareh Sarrafzadeh.
\newblock Prototypical human-ai collaboration behaviors from llm-assisted
  writing in the wild.
\newblock In {\em Proceedings of the 2025 Conference on Empirical Methods in
  Natural Language Processing}, pages 16830--16857, 2025.

\bibitem[MLT{\etalchar{+}}24]{maharana2024evaluating}
Adyasha Maharana, Dong-Ho Lee, Sergey Tulyakov, Mohit Bansal, Francesco
  Barbieri, and Yuwei Fang.
\newblock Evaluating very long-term conversational memory of llm agents.
\newblock In {\em Proceedings of the 62nd Annual Meeting of the Association for
  Computational Linguistics (Volume 1: Long Papers)}, pages 13851--13870, 2024.

\bibitem[MMHZ20]{min2020ambigqa}
Sewon Min, Julian Michael, Hannaneh Hajishirzi, and Luke Zettlemoyer.
\newblock Ambigqa: Answering ambiguous open-domain questions.
\newblock In {\em Proceedings of the 2020 conference on empirical methods in
  natural language processing (EMNLP)}, pages 5783--5797, 2020.

\bibitem[MSW{\etalchar{+}}23]{mu2023clarifygpt}
Fangwen Mu, Lin Shi, Song Wang, Zhuohao Yu, Binquan Zhang, Chenxue Wang,
  Shichao Liu, and Qing Wang.
\newblock Clarifygpt: Empowering llm-based code generation with intention
  clarification.
\newblock {\em arXiv preprint arXiv:2310.10996}, 2023.

\bibitem[PGA{\etalchar{+}}25]{park2025know}
Young-Jin Park, Kristjan Greenewald, Kaveh Alim, Hao Wang, and Navid Azizan.
\newblock Know what you don't know: Uncertainty calibration of process reward
  models.
\newblock {\em arXiv preprint arXiv:2506.09338}, 2025.

\bibitem[POC{\etalchar{+}}23]{park2023generative}
Joon~Sung Park, Joseph O'Brien, Carrie~Jun Cai, Meredith~Ringel Morris, Percy
  Liang, and Michael~S Bernstein.
\newblock Generative agents: Interactive simulacra of human behavior.
\newblock In {\em Proceedings of the 36th annual acm symposium on user
  interface software and technology}, pages 1--22, 2023.

\bibitem[PWJ{\etalchar{+}}25]{pan2025memory}
Zhuoshi Pan, Qianhui Wu, Huiqiang Jiang, Xufang Luo, Hao Cheng, Dongsheng Li,
  Yuqing Yang, Chin-Yew Lin, H~Vicky Zhao, Lili Qiu, et~al.
\newblock On memory construction and retrieval for personalized conversational
  agents.
\newblock {\em arXiv preprint arXiv:2502.05589}, 2025.

\bibitem[QLP{\etalchar{+}}25]{qian2025userbench}
Cheng Qian, Zuxin Liu, Akshara Prabhakar, Zhiwei Liu, Jianguo Zhang, Haolin
  Chen, Heng Ji, Weiran Yao, Shelby Heinecke, Silvio Savarese, et~al.
\newblock Userbench: An interactive gym environment for user-centric agents.
\newblock {\em arXiv preprint arXiv:2507.22034}, 2025.

\bibitem[QSA{\etalchar{+}}26]{qiu2026bayesian}
Linlu Qiu, Fei Sha, Kelsey Allen, Yoon Kim, Tal Linzen, and Sjoerd van
  Steenkiste.
\newblock Bayesian teaching enables probabilistic reasoning in large language
  models.
\newblock {\em Nature Communications}, 2026.

\bibitem[Rit18]{ritchey2018general}
Tom Ritchey.
\newblock General morphological analysis as a basic scientific modelling
  method.
\newblock {\em Technological Forecasting and Social Change}, 126:81--91, 2018.

\bibitem[RJL18]{rajpurkar2018know}
Pranav Rajpurkar, Robin Jia, and Percy Liang.
\newblock Know what you don’t know: Unanswerable questions for squad.
\newblock In {\em Proceedings of the 56th Annual Meeting of the Association for
  Computational Linguistics (Volume 2: Short Papers)}, pages 784--789, 2018.

\bibitem[RWF{\etalchar{+}}23]{rahmani2023survey}
Hossein~A Rahmani, Xi~Wang, Yue Feng, Qiang Zhang, Emine Yilmaz, and Aldo
  Lipani.
\newblock A survey on asking clarification questions datasets in conversational
  systems.
\newblock In {\em Proceedings of the 61st Annual Meeting of the Association for
  Computational Linguistics (Volume 1: Long Papers)}, pages 2698--2716, 2023.

\bibitem[SBL{\etalchar{+}}18]{saeidi2018interpretation}
Marzieh Saeidi, Max Bartolo, Patrick Lewis, Sameer Singh, Tim Rockt{\"a}schel,
  Mike Sheldon, Guillaume Bouchard, and Sebastian Riedel.
\newblock Interpretation of natural language rules in conversational machine
  reading.
\newblock In {\em Proceedings of the 2018 Conference on Empirical Methods in
  Natural Language Processing}, pages 2087--2097, 2018.

\bibitem[SSZ25]{song2025hallucination}
Linxin Song, Taiwei Shi, and Jieyu Zhao.
\newblock The hallucination tax of reinforcement finetuning.
\newblock {\em arXiv preprint arXiv:2505.13988}, 2025.

\bibitem[SVW{\etalchar{+}}24]{singh2024personal}
Harmanpreet Singh, Nikhil Verma, Yixiao Wang, Manasa Bharadwaj, Homa Fashandi,
  Kevin Ferreira, and Chul Lee.
\newblock Personal large language model agents: A case study on tailored travel
  planning.
\newblock In {\em Proceedings of the 2024 conference on empirical methods in
  natural language processing: industry track}, pages 486--514, 2024.

\bibitem[TMZ{\etalchar{+}}23]{tian2023just}
Katherine Tian, Eric Mitchell, Allan Zhou, Archit Sharma, Rafael Rafailov,
  Huaxiu Yao, Chelsea Finn, and Christopher~D Manning.
\newblock Just ask for calibration: Strategies for eliciting calibrated
  confidence scores from language models fine-tuned with human feedback.
\newblock In {\em Proceedings of the 2023 Conference on Empirical Methods in
  Natural Language Processing}, pages 5433--5442, 2023.

\bibitem[TYH{\etalchar{+}}25]{tan2025prospect}
Zhen Tan, Jun Yan, I-Hung Hsu, Rujun Han, Zifeng Wang, Long Le, Yiwen Song,
  Yanfei Chen, Hamid Palangi, George Lee, et~al.
\newblock In prospect and retrospect: Reflective memory management for
  long-term personalized dialogue agents.
\newblock In {\em Proceedings of the 63rd Annual Meeting of the Association for
  Computational Linguistics (Volume 1: Long Papers)}, pages 8416--8439, 2025.

\bibitem[VZY{\etalchar{+}}26]{vijayvargiya2026ambig}
Sanidhya Vijayvargiya, Xuhui Zhou, Akhila Yerukola, Maarten Sap, and Graham
  Neubig.
\newblock Ambig-swe: Interactive agents to overcome underspecificity in
  software engineering.
\newblock In {\em The Fourteenth International Conference on Learning
  Representations}, 2026.

\bibitem[WGP{\etalchar{+}}25]{wu2025collabllm}
Shirley Wu, Michel Galley, Baolin Peng, Hao Cheng, Gavin Li, Yao Dou, Weixin
  Cai, James Zou, Jure Leskovec, and Jianfeng Gao.
\newblock Collabllm: From passive responders to active collaborators.
\newblock {\em arXiv preprint arXiv:2502.00640}, 2025.

\bibitem[WWY{\etalchar{+}}24]{wu2024longmemeval}
Di~Wu, Hongwei Wang, Wenhao Yu, Yuwei Zhang, Kai-Wei Chang, and Dong Yu.
\newblock Longmemeval: Benchmarking chat assistants on long-term interactive
  memory.
\newblock {\em arXiv preprint arXiv:2410.10813}, 2024.

\bibitem[WZZN25]{wang2025adaptive}
Jimmy Wang, Thomas Zollo, Richard Zemel, and Hongseok Namkoong.
\newblock Adaptive elicitation of latent information using natural language.
\newblock {\em arXiv preprint arXiv:2504.04204}, 2025.

\bibitem[XLM{\etalchar{+}}25]{xu2025mem}
Wujiang Xu, Zujie Liang, Kai Mei, Hang Gao, Juntao Tan, and Yongfeng Zhang.
\newblock A-mem: Agentic memory for llm agents.
\newblock {\em arXiv preprint arXiv:2502.12110}, 2025.

\bibitem[ZGG{\etalchar{+}}24]{zhong2024memorybank}
Wanjun Zhong, Lianghong Guo, Qiqi Gao, He~Ye, and Yanlin Wang.
\newblock Memorybank: Enhancing large language models with long-term memory.
\newblock In {\em Proceedings of the AAAI conference on artificial
  intelligence}, volume~38, pages 19724--19731, 2024.

\bibitem[ZHU{\etalchar{+}}25]{zhang2025agentic}
Qizheng Zhang, Changran Hu, Shubhangi Upasani, Boyuan Ma, Fenglu Hong,
  Vamsidhar Kamanuru, Jay Rainton, Chen Wu, Mengmeng Ji, Hanchen Li, et~al.
\newblock Agentic context engineering: Evolving contexts for self-improving
  language models.
\newblock {\em arXiv preprint arXiv:2510.04618}, 2025.

\bibitem[Zwi67]{zwicky1967morphological}
Fritz Zwicky.
\newblock The morphological approach to discovery, invention, research and
  construction.
\newblock In {\em New methods of thought and procedure: contributions to the
  symposium on methodologies}, pages 273--297. Springer, 1967.

\bibitem[ZXW{\etalchar{+}}23]{zhu2023calibration}
Chiwei Zhu, Benfeng Xu, Quan Wang, Yongdong Zhang, and Zhendong Mao.
\newblock On the calibration of large language models and alignment.
\newblock In {\em Findings of the Association for Computational Linguistics:
  EMNLP 2023}, pages 9778--9795, 2023.

\end{thebibliography}
